\documentclass[pdflatex]{sn-jnl}% Math and Physical Sciences Numbered Reference Style
\usepackage{graphicx}%
\usepackage{multirow}%
\usepackage{amsmath,amssymb,amsfonts}%
\usepackage{amsthm}% 
\usepackage{mathrsfs}%
\usepackage[title]{appendix}%
\usepackage{xcolor}%
\usepackage{textcomp}%
\usepackage{manyfoot}%
\usepackage{booktabs}%
\usepackage{algorithm}%
\usepackage{algorithmicx}%
\usepackage{algpseudocode}%
\usepackage{listings}%
\usepackage{mathtools}
\usepackage[authoryear]{natbib}

\theoremstyle{thmstyleone}

\theoremstyle{thmstyletwo}

\theoremstyle{thmstylethree}

\usepackage[object=vectorian]{pgfornament}
\usepackage[mathlines]{lineno}

\newcommand{\hlf}{\frac{1}{2}}
\newcommand{\lt}{\left}
\newcommand{\rt}{\right}

\newcommand{\tbf}[1]{\textbf{#1}}

\newcommand{\erm}{{\rm e}}
\begin{document}

\title[An angular-momentum preserving dissipative model for the point-mass $N$-body problem]{An angular-momentum preserving dissipative model for the point-mass $N$-body problem}


% region names_authors
\author*[1]{\fnm{Matheus} \sur{Lazarotto}}\email{m.j.lazarotto@gmail.com}

\author[1]{\fnm{Clodoaldo} \sur{Ragazzo}}\email{ragazzo@usp.br}

\affil*[1]{\orgdiv{Instituto de Matemática, Estatística e Ciências da Computação}, \orgname{Universidade de São Paulo}, \orgaddress{\street{}\city{São Paulo}, \postcode{05508-090}, \state{SP}, \country{Brasil}}}

% \affil[2]{\orgdiv{Department}, \orgname{Organization}, \orgaddress{\street{Street}, \city{City}, \postcode{10587}, \state{State}, \country{Country}}}
% endregion

% region abstract
\abstract{
A simple mathematical model emulating energy dissipation due to tidal effects is proposed.
In this model, forces acting between masses remove energy but preserve the total angular momentum of the system.
We study the effect of such forces on the particular family of orbits in central configurations, and show that a specific dependence 
on the mutual distances between the bodies leads to homographic equations equivalent to those of the two-body problem with dissipation.
We then describe in detail the topology of solutions of the dissipative two-body system via Poincaré compactification. Finally, we present equations averaged over Keplerian motion showing no influence of the dissipation on periapsis precession.
}
% endregion

\keywords{Celestial Mechanics, $N$-body problem, Dissipative systems}

\maketitle

\section{Introduction}\label{sec:introduction}

When predicting the fate of $N$-bodies mutually attracted by gravity, two main frameworks are often 
used to model their dynamical evolution. 
In the conservative one, orbits are either regular (periodic or quasi-periodic) or chaotic, 
where bodies may escape, collide, or revolve around each other indefinitely \citep{Smale,Contopoulosbook,Ferraz-Mellobook}.
In the dissipative one, these scenarios can still occur, but pairs of bounded bodies, given sufficient time, 
eventually end up in spin-orbit coupling, facing each other while moving in  
circular orbits, or collide \citep{Darwin}. 

In fact, the spin-orbit coupling among bounded pairs extends to systems with any number of bodies 
($N > 2$), as the whole system is progessively reduced, pair by pair, to one of the final states described: 
escapes, collisions, or synchronizations \citep{Scheeres2012}.

In detail, one can start by assuming the conservative dynamics for $N$-bodies, 
generally taking place on the hypersurface $E: \mathbb{R}^{3N} \times T\mathbb{R}^{3N} \to \mathbb{R}$, 
corresponding to the system's mechanical energy

\begin{equation}\label{eq:energy}
    E(\tbf{r}, \dot{\tbf{r}}) = T(\dot{\tbf{r}}) + U(\tbf{r}),
\end{equation}
\\\noindent
for $\tbf{r} = \lt(\tbf{r}_1 \; ... \; \tbf{r}_N\rt) = \lt(x_1 \; y_1 \; z_1 \; ... \; x_N \; y_N \; z_N\rt) \in \mathbb{R}^{3N}$ 
as the cartesian position for all $N$-bodies, and $\dot{\tbf{r}} = \frac{d \tbf{r}}{dt} \in T\mathbb{R}^{3N}$ as their velocities. 
The kinetic energy

\begin{equation}\label{eq:kinetic}
    % T(\dot{\tbf{r}}) = \hlf \sum_{i=1}^N \int_{\mathcal{B}_i} \lt(\dot{\tbf{r}} \cdot \dot{\tbf{r}}\rt) dm_i (\tbf{r})
    T(\dot{\tbf{r}}) = \hlf \sum_{i=1}^N m_i \lt(\dot{\tbf{r}}_i \cdot \dot{\tbf{r}}_i\rt)
\end{equation}
\\\noindent
and the gravitational potential energy

\begin{equation}\label{eq:potential}
    % U(\tbf{r}) = -G \sum_{i=1}^{N-1} \sum_{j=i+1}^{N} \int_{\mathcal{B}_i} \int_{\mathcal{B}_j} \frac{dm_i dm_j}{|\tbf{r}_{ij}|},
    U(\tbf{r}) = -G \sum_{i=1}^{N-1} \sum_{j=i+1}^{N} \frac{m_i m_j}{|\tbf{r}_{ij}|},
\end{equation}
\\\noindent
are computed for masses $m_i$
% elements $dm_i(\tbf{r}) = \rho_i(\tbf{r})dV$ for extensive bodies defined by the volume sets $\mathcal{B}(\tbf{r}) = \{\mathcal{B}_i\}_{i=1}^{N}$, 
with relative distances $\tbf{r}_{ij} = \tbf{r}_i - \tbf{r}_j \in \mathbb{R}^3$. 
% For point-mass distributions, $dm_i (\tbf{r}) = m_i \delta(|\tbf{r} - \tbf{r}_i|)$, which reduces to the classical particle system.

If no external forces act on the system, and regardless of internal constraints, the dynamics will 
preserve both the energy function (\ref{eq:energy}) and the total angular momentum 

\begin{equation}\label{eq:angular-mom}
    % \tbf{L} = \sum_{i=1}^{N} \int_{\mathcal{B}_i} \lt(\tbf{r} \times \dot{\tbf{r}}\rt) dm_i(\tbf{r}),
    \tbf{L} = \sum_{i=1}^{N} m_i \lt(\tbf{r}_i \times \dot{\tbf{r}}_i\rt),
\end{equation}
\\\noindent
due to the central nature of the forces from the gravitational potential (\ref{eq:potential}).
The expressions (\ref{eq:energy}) to (\ref{eq:angular-mom}) hold for a reference frame fixed 
at and moving with the center of mass of the whole system, satisfying

\begin{equation}\label{eq:reference-frame}
    % \sum_{i=1}^N \int_{\mathcal{B}_i} \tbf{r} \, dm_i(\tbf{r}) = 
    % \sum_{i=1}^N \int_{\mathcal{B}_i} \dot{\tbf{r}} \, dm_i(\tbf{r}) = 0.
    \sum_{i=1}^N m_i \tbf{r}_i = \sum_{i=1}^N m_i \dot{\tbf{r}}_i  = 0.
\end{equation}
\\\indent
In this context, \citet{Smale} proved that for any gravitational 
system of point-masses, the critical points of the energy function \eqref{eq:energy} 
restricted to a constant angular momentum hypersurface \eqref{eq:angular-mom}
correspond to relative equilibrium configurations, i.e.,

\begin{equation}
    \frac{d |\tbf{r}_{ij}|}{dt} = 0 \qquad \forall i,j.
\end{equation}
\\\indent
% One can picture such configurations in any movement of the system that rotates as a solid body, with 
% fixed mutual distances.

On top of that, \citet{Moeckel1990} showed that for $N > 2$ any relative equilibrium 
is a saddle-point of the energy function, when restricted to a constant angular momentum level. 
Later, \citet{Moeckel2017} proved similar results for systems of interacting rigid bodies.

% Although these results hold in the conservative scenario, weak sources of dissipation are always present 
% in celestial systems and become significant over long timescales.
% Nonetheless, these two fundamental results from the conservative limit still enable predictions about the 
% fate of many-body systems 
In the presence of tides, which dissipate energy but not angular momentum, a distinct conclusion 
can be derived from the results of Smale and Moeckel.
For $N \geq 2$, energy dissipation leads to two possible physical scenarios: 
either bodies collide and/or move apart and ultimately become isolated 
masses; or, after successive collisions and escapes, pairs of masses remain in circular motion, 
where no further energy can be dissipated (see \citep{Scheeres2002,Scheeres2012} for details).
% In other words, the dynamics in phase space slowly decays along the energy surface while constrained 
% to a angular momentum surface level. 
% Energy may decay toward a positive value for single bodies, or toward a global negative minimum 
% that exists only for two bodies in relative equilibrium. 
The combination of Smale's and Moeckel's results assert that, despite the existence of 
relative equilibria, their instability prevents systems with more than two bodies to remain 
in such states indefinitely.

It is well known that the behavior observed in nature is dissipative, 
as different mechanisms act on any group of celestial bodies due to 
internal friction (tidal effects), several types of drag (motion within 
gases, dust, plasma, etc.), and radiative effects \citep{Ferraz-Mellobook}.
% tidal effects, internal friction, surface interactions, solar wind drag, Poyting-Roberston 
% drag, among others \citep{Ferraz-Mellobook}.
However, in empty space, these dissipative forces are generally weak, and 
the timescales required to reach final states are often unattainable.
As a result, one typically observes systems in transient configurations lasting for millions 
or even billions of years, closely resembling stability. 
This sugests that the nature of dissipative effects allows for prolonged 
transients near semi-equilibrium states, making them relevant to the dynamical evolution 
as such states likely represent most of a system's lifetime and determine their path 
toward any final state it may reach.

Several models for dissipative effects have been studied 
\citep{Gabriel,Wang,Celletti2007,Celletti2011,Margheri,Ferraz-Mello}, 
mainly involving Poynting-Robertson drag, solar wind, atmospheric drag, or surface contact.
In these cases, dissipation typically acts as a drag force opposing the body's velocity through 
a damping term $-\alpha \dot{\tbf{r}}$ in the equation of motion

\begin{equation}
      m\ddot{\tbf{r}} = -\alpha \dot{\tbf{r}} - \nabla U(\tbf{r}).
\end{equation}
\\\indent
Even though these models correspond to relevant and realistic scenarios, they remove both energy 
and angular momentum from the system.
By contrast, at planetary scale, the most prominent dissipative effect is gravitational in origin, 
arising from tides, which preserves angular momentum \citep{Goldreich,Meyer}.

% For instance, the power dissipated by tide forces in a satellite of radius $R$,  
% orbiting a planet of mass $m$ in an eccentric orbit with semi-major axis $a$ is roughly 
% $G m^2 R^5 / (a^6 \  n \ h_i)$, with $G$ being the gravitational constant, $n$ the orbital 
% frequency and $h_i$ a constant that depends on the orbit eccentricity and on the physical 
% properties of the body (the imaginary part of the Love number). 
% This implies a dissipation of a different nature, one that dissipates energy while conserving 
% angular momentum.

In this context, we introduce a simple model for dissipation between celestial bodies 
that, while dissipating energy, preserves angular momentum. 
This is achieved by introducing a central damping force that breaks time symmetry while 
maintaining rotational symmetry.

We remark that dissipative forces related to, although distinct from, the one proposed here have previously 
appeared in works by C. Ai, M. Margheri, M. Misquero, F. Diacu, S. Breiter, and A. Jackson.

\cite{Margheri2020} considered a dissipative force with a more general radial dependence than ours. 
Their force, however, does not preserve angular momentum. Consequently, their model exhibits qualitative 
behavior different from that obtained here, although parts of their analysis are closely related to ours.

\cite{Diacu1999} and \cite{Breiter1998} studied generalized Stokes drag forces decomposed into radial and 
tangential components. Their formulations include purely radial dissipation as the particular case 
$\beta=0$ in their notation, which is related to the type of force considered here. 
This particular case, however, was not investigated separately in those works. 
Moreover, these models do not, in general, preserve angular momentum and therefore do not 
share some of the qualitative properties of our system.

In his PhD thesis, written under the supervision of Mark Levi, \cite{Chong} considered a dissipative force 
that decreases the mechanical energy while preserving angular momentum. The dependence of Ai's force on the 
mutual distance is, however, substantially different from that of the force proposed here. 
Nevertheless, the preservation of angular momentum leads to some qualitative features common to both models, 
such as the circularization of orbits in the dissipative Kepler problem. 
These features are absent from the models of Margheri and Misquero, Diacu, and Breiter and Jackson. 
Ai also studied dissipative effects in the restricted three-body problem, a direction that we are currently 
pursuing.

The force introduced here is therefore related to these earlier models through some of its dissipative and 
conservation properties, but it is distinct from all of them. 
Its remarkable feature is that, under dissipation, it leaves invariant the manifold that, in the conservative 
problem, is formed by the homographic solutions. 
This invariance does not follow merely from the preservation of angular momentum; it depends on the particular 
form of the force considered here. 
To the best of our knowledge, this specific force and its associated invariance property have not previously 
appeared in the literature.

In what follows, we present the model and its main properties in Section \ref{sec:model}.
Section \ref{sec:central-configs} discusses its effects on planar central configurations, 
while Section \ref{sec:two-bodies} and \ref{sec:averaging} provide topological and 
numerical results for the simpler case of two bodies -- results that also apply to planar 
central motions for any number of bodies.

\section{Model}\label{sec:model}

To incorporate energy dissipation while conserving angular momentum, 
we draw on Mignard's tidal force approximation for two nearly spherical 
bodies (eq. (5) in \citep{Mignard}). 
We simplify his expression to a point-mass scenario by neglecting rotational terms, 
yielding the dissipative force $\tbf{f}^{\,ij}_\textrm{diss}$ between a pair 
of masses $m_i$ and $m_j$
\begin{equation}\label{eq:dissipative-force}
      \tbf{f}^{\,ij}_\textrm{diss}(\dot{\tbf{r}}_{ij},\tbf{r}_{ij}) = -\alpha \frac{G m_i m_j}{|\tbf{r}_{ij}|^d} \lt(\dot{\tbf{r}}_{ij} \cdot \hat{\tbf{r}}_{ij}\rt) \hat{\tbf{r}}_{ij},
\end{equation}
\noindent
where $\hat{\tbf{r}}_{ij} = \frac{\tbf{r}_i - \tbf{r}_j}{|\tbf{r}_i - \tbf{r}_j|} \in \mathbb{R}^3$ is the 
versor connecting the $i$-th and $j$-th bodies.
The dissipation strength $\alpha$ is constant and has units of $\left[ L^{d-3} T \right]$. 
Physically, it reflects rheological and viscoelastic effects\footnote{\label{foot1}
In principle, the dissipation constant $\alpha$ may vary from body to body, so that
\[
\tbf{f}^{\,ij}_\textrm{diss}(\dot{\tbf{r}}_{ij},\tbf{r}_{ij}) = -\alpha_i\alpha_j \frac{G m_i m_j}{|\tbf{r}_{ij}|^d} \lt(\dot{\tbf{r}}_{ij} \cdot \hat{\tbf{r}}_{ij}\rt) \hat{\tbf{r}}_{ij},\] 
In a forthcoming paper, focusing on the three-body problem, we use this formulation. Here we assume a uniform $\alpha$ for all bodies. This distinction is irrelevant in the case of two bodies.
}.
Since the current study focuses on a mathematical approach, we treat 
the radial dependence exponent $d$ as a free parameter, as it will be discussed 
further in the text. 
A direct comparison to Mignard's model would yield $d=8$.

The force (\ref{eq:dissipative-force}) is pairwise anti-symmetric ($\tbf{f}_{ij} = -\tbf{f}_{ji}$), 
dissipative because it depends on the relative velocity $(\dot{\tbf{r}}_{ij} = \dot{\tbf{r}}_i - \dot{\tbf{r}}_j)$, 
and central ($\tbf{f}_{ij} \, \| \,\hat{\tbf{r}}_{ij}$).
It is also homogeneous in the positions and velocities
$\tbf{f}^{\,ij}_\textrm{diss}(\nu \dot{\tbf{r}}_{ij}, \mu \tbf{r}_{ij}) = \nu \mu^{-d} \tbf{f}^{\,ij}_\textrm{diss}(\dot{\tbf{r}}_{ij}, \tbf{r}_{ij})$.

We start by considering the conservative Lagrangian for $N$ point masses

\begin{equation}\label{eq:lagrangian}
    \mathcal{L} = \hlf \sum_{i=1}^{N} m_i \lt(\dot{\tbf{r}}_i \cdot \dot{\tbf{r}}_i\rt) 
                + G \sum_{i=1}^{N-1} \sum_{j=i+1}^{N} \frac{m_i m_j}{|\tbf{r}_{ij}|},
\end{equation}
\\\noindent
and introduce dissipation via the Rayleigh function

\begin{equation}\label{eq:rayleigh}
    \mathcal{D} = \frac{\alpha}{2} \sum_{i=1}^{N-1} \sum_{j=i+1}^{N} \frac{G m_i m_j}{|\tbf{r}_{ij}|^{d}} \lt(\dot{\tbf{r}}_{ij} \cdot \hat{\tbf{r}}_{ij}\rt)^2,
\end{equation}
\\\noindent
which modify the Euler-Lagrange equations as

\begin{equation}\label{eq:euler-lagrange}
    \frac{d}{dt}\lt(\nabla_{\dot{\tbf{r}}_i} \mathcal{L}\rt) = \nabla_{\tbf{r}_i} \mathcal{L} - \nabla_{\dot{\tbf{r}}_i} \mathcal{D};
\end{equation}
\\\noindent
yielding the equations of motion

\begin{equation}\label{eq:motion-equation}
      m_i \ddot{\tbf{r}}_i = - \sum_{j \neq i}^{N} \frac{G m_i m_j}{|\tbf{r}_{ij}|^3} \, \tbf{r}_{ij}
                             + \sum_{j \neq i}^{N} \tbf{f}^{\,ij}_\textrm{diss}(\dot{\tbf{r}}_{ij},\tbf{r}_{ij}) ,
\end{equation}
for $\tbf{r}_i \in \mathbb{R}^3$ as the position of each mass $m_i$ relative to the center of mass.

\subsection{Energy dissipation}\label{subsec:energy-dissipation}

To assess how the dissipative force (\ref{eq:dissipative-force}) affects the system, 
we examine two conservative invariants of motion, namely the energy and angular momentum.
Because the force is central, it immediately follows that the angular momentum 
(\ref{eq:angular-mom}) will remain conserved.

As for the energy, we can evaluate its dissipation by taking the scalar product of the 
equation of motion (\ref{eq:motion-equation}) with the velocity $\dot{\tbf{r}}$ 
\begin{equation}
\begin{split}
    M \ddot{\tbf{r}} \cdot \dot{\tbf{r}} + \nabla_{\tbf{r}} U(\tbf{r}) \cdot \dot{\tbf{r}} &= -2 \mathcal{D}(\dot{\tbf{r}}, \tbf{r})
\end{split}
\end{equation}
\noindent
and identifying the left hand side as the derivative of the total energy $E = T + U$, 
and therefore
\begin{equation}
    \frac{dE}{dt} = -2 \mathcal{D} \leq 0.
\end{equation}
\\\indent
Dissipation will only cease when $\mathcal{D} = 0$, which exclusively 
occurs in configurations of relative equilibria, or rotational central configurations, 
where the whole system rotates with all pairs $\tbf{r}_{ij}$ fixed.

% In fact, the only case where dissipation can cease indefinitely is for two bodies. 
% For $N=2$, energy reaches its minimum at the circular relative equilibrium and 
% remains there.
% Whenever other bodies are present, even if the movement is locally circular, 
% the presence of any remaining body keeps $\mathcal{D} \neq 0$ at all times, 
% unless they all rotate as a rigid body.

\section{Central configurations}\label{sec:central-configs}

% region central-configs-intro
In this section we introduce some necessary concepts on central configurations (CCs) 
and analyze the effect of the dissipative force (\ref{eq:dissipative-force}) on 
this family of orbits when restricted to the planar case.
For further read on CCs, we direct the reader to Moeckel's work 
(Chap. 2 in \citep{Moeckelbook}) on which we base our calculations.

Given a set of $N$-bodies with masses $m_1, \dots, m_N$, a CC is an arrangement 
where the force on each body points towards the center of mass of the system, 
that is

\begin{equation}
    \nabla U (\tbf{r}) + \lambda \tbf{M} \lt(\tbf{r} - \tbf{c}\rt) = 0,
\end{equation}
\\\noindent
for some constant $\lambda \in \mathbb{R}$. Here $\tbf{r} \in \mathbb{R}^{3N}$ (or $\mathbb{R}^{2N}$) 
is the global state vector with all position coordinates for the $N$-bodies.
The mass matrix 

\begin{equation*}\label{eq:mass-matrix}
    \tbf{M} = \text{diag}_N\lt(\tbf{M}_i\rt)
            = \begin{pmatrix}
                \mathbf{M}_1 & \mathbf{0} & \cdots & \mathbf{0} \\
                \mathbf{0} & \mathbf{M}_2 & \cdots & \mathbf{0} \\
                \vdots & \vdots & \ddots & \vdots \\
                \mathbf{0} & \mathbf{0} & \cdots & \mathbf{M}_N
              \end{pmatrix} 
            \in \mathbb{M}_{3N} \; (\text{or}\;\mathbb{M}_{2N}),
\end{equation*}
\\\noindent
stores the masses for each body, with $\tbf{M}_i = m_i \mathbf{I}_3$ (or $m_i \mathbf{I}_2$), for $i = 1, ..., N$, and 
where $\mathbf{I}_3$ ($\mathbf{I}_2$) is the identity matrix in $\mathbb{R}^3$ ($\mathbb{R}^2$). 
The vector $\tbf{c} \in \mathbb{R}^{3N}$ is a stacked vector of the center of mass $\tbf{c}_0$, such that

\begin{equation*}\label{eq:state-center-of-mass}
    \lt(\tbf{r} - \tbf{c}\rt) = \lt(\tbf{r}_1 - \tbf{c}_0 \;\; ... \;\; \tbf{r}_n - \tbf{c}_0\rt),
\end{equation*}
\noindent
with 
\begin{equation*}\label{eq:local-center-of-mass}
    \tbf{c}_0 = \frac{1}{M}\sum_{i=1}^{N} m_i \tbf{r}_i
\end{equation*}
\\\noindent
as the center of mass in 3D (or 2D); $M = \sum_{i=1}^{N} m_i$ is the total mass. 
Normally $\tbf{c}_0$ can be made zero if the coordinate reference frame is centered 
at it.

It can be shown that $\lambda = \frac{U(\tbf{r})}{I(\tbf{r})}$, where $I(\tbf{r})$ is the moment 
of inertia of the system

\begin{equation}\label{eq:moment-inertia}
    I(\tbf{r}) = \lt(\tbf{r} - \tbf{c}\rt)^T \tbf{M} \lt(\tbf{r} - \tbf{c}\rt) 
               = \sum_{i=1}^{N} m_i \lt|\tbf{r}_i - \tbf{c}_0\rt|^2 
               = \frac{1}{M} \sum_{i=1}^{N} \sum_{j < i}^{N} m_i m_j r_{ij}^2.
\end{equation}
% endregion

\subsection{Planar case}\label{sec:planar-central-configs}

When restricted to a plane, the motion of any CC can be written in homographic form, 
combining a scaling of the configuration (homothetic motion) and a 
rotation (rigid body motion)

\begin{equation}\label{eq:homographic}
    \tbf{r}(t) - \tbf{c}(t) = s(t) \tbf{Q}(\theta(t)) \lt(\tbf{r}_0 - \tbf{c}_0\rt)
\end{equation}
\\\noindent
where $s(t)\in\mathbb{R}$ is the scaling factor, and $\tbf{Q}(\theta(t))$ is a rotation matrix 
about the axis perpendicular to the plane, rotating each body uniformly by $\theta(t)$

\begin{equation}\label{eq:planar-rot-matrix}
    \tbf{Q}(\theta(t)) = \textrm{diag}_N \lt(\tbf{q}(\theta(t))\rt) \in \mathbb{M}_{2N}
\end{equation}
\noindent
where

\begin{equation}\label{eq:rotation-planar}
    \tbf{q}(\theta(t)) = \begin{pmatrix} \cos(\theta) & -\sin(\theta) \\
                                         \sin(\theta) & \cos(\theta) 
                         \end{pmatrix} \in \textrm{SO}(2);
\end{equation}
\\\noindent
$(\tbf{r}_0 - \tbf{c}_0) \in \mathbb{R}^{2N}$ is the initial state of the CC. 
Note that $\tbf{c}_0 \in \mathbb{R}^{2N}$ here generates an ambiguity with the vector 
$\tbf{c}_0 \in \mathbb{R}^{2}$, denoting the position of the center of mass in the plane of motion. 
Although related, the ambiguity should not bring any conflict in the following calculations.

Over a homographic configuration ($\tbf{r}_h,\dot{\tbf{r}}_h$) satisfying (\ref{eq:homographic}) 
the Rayleigh dissipation function (\ref{eq:rayleigh}) can be evaluated as
\begin{equation}\label{eq:rayleigh-homographic}
\begin{split}
    \nabla_{\dot{\tbf{r}}_i} \mathcal{D}(\tbf{r}_h,\dot{\tbf{r}}_h) 
    &= \sum_{j \neq i}^{N} \mu_{ij} s^{-d} D(r_{ij, 0}) 
      \lt\{\dot{s} \lt(\tbf{q} \tbf{r}_{ij,0}\rt) \cdot \lt(\tbf{q} \hat{\tbf{r}}_{ij,0}\rt) 
                  + s \dot{\theta} \lt(\frac{d \tbf{q}}{d\theta} \tbf{r}_{ij,0}\rt) \cdot \lt(\tbf{q} \hat{\tbf{r}}_{ij,0}\rt) \rt\} 
      \lt(\tbf{q} \hat{\tbf{r}}_{ij,0}\rt) \\
    \nabla_{\dot{\tbf{r}}_i} \mathcal{D}(\tbf{r}_h,\dot{\tbf{r}}_h)
    &= \dot{s} s^{-d} \tbf{q}  \lt(\sum_{j \neq i}^{N} \mu_{ij} D(r_{ij, 0}) \tbf{r}_{ij,0}\rt) \\
\end{split}
\end{equation}
where $\mu_{ij} = \alpha G m_i m_j$ and $D(r_{ij, 0}) = |\tbf{r}_{ij, 0}|^{-d}$. 
When replacing the homographic form (\ref{eq:homographic}) into (\ref{eq:rayleigh}), 
the following identities were used
\begin{equation}\label{eq:cc-first-dt}
\begin{split}
    &\frac{d \lt(\tbf{r} - \tbf{c}\rt)}{dt} = \lt(\dot{s}(t) \tbf{Q}(\theta(t)) + s(t) \tbf{Q}' \dot{\theta}(t)\rt) \lt(\tbf{r}_0 - \tbf{c}_0\rt) \\
    &\frac{d^2 \lt(\tbf{r} - \tbf{c}\rt)}{dt^2} = \lt(\ddot{s}(t) \tbf{Q}(\theta(t)) + 2 \dot{s}(t) \tbf{Q}' \dot{\theta}(t) + s(t) \tbf{Q}'' \dot{\theta}^2(t) + s(t) \tbf{Q}' \ddot{\theta}(t)\rt) \lt(\tbf{r}_0 - \tbf{c}_0\rt)
\end{split}
\end{equation}
where
\begin{equation*}\label{eq:rot-first-dtheta}
    \tbf{Q}' = \frac{d \tbf{Q}(\theta)}{d\theta} \;\to\; 
    \frac{d\tbf{q}}{d\theta} 
    % = \begin{pmatrix} -\sin(\theta) & -\cos(\theta) \\
    %                    \cos(\theta) & -\sin(\theta) \end{pmatrix} = 
    = \textbf{j} \textbf{q}
    \qquad\text{with}\qquad \tbf{j} = \tbf{q}(\pi/2),
\end{equation*}
and
\begin{equation*}\label{eq:rot-second-dtheta}
    \tbf{Q}'' = \frac{d^2 \tbf{Q}(\theta)}{d\theta^2} \to \frac{d^2\tbf{q}}{d\theta^2} 
    = -\tbf{q}.
\end{equation*}
\\\noindent
It was also used that rotations preserve inner products and that, when
restricted to the plane, they commute, implying 
\begin{equation*}\label{eq:orthogonality}
    \lt(\tbf{j} \tbf{q} \tbf{r}_{ij,0}\rt) \cdot \lt(\tbf{q} \hat{\tbf{r}}_{ij,0}\rt) = 
    \lt(\tbf{j} \tbf{r}_{ij,0}\rt) \cdot \hat{\tbf{r}}_{ij,0} = 0.
\end{equation*}

\subsection{The special case $d=3$}

For $d=3$, the dissipative force $\nabla_{\dot{\tbf{r}}_i} \mathcal{D}$ in 
(\ref{eq:rayleigh-homographic}) can be expressed in terms of the gradient of 
the gravitational potential
\begin{equation*}
\begin{split}
    % &\nabla_{\dot{\tbf{r}}_i} \mathcal{D}(\tbf{r}_h,\dot{\tbf{r}}_h) = \alpha \dot{s} s^{-3} \tbf{q} \lt(\sum_{j \neq i}^{N} G m_i m_j r_{ij,0}^{-2} \hat{\tbf{r}}_{ij,0}\rt) \\
    &\nabla_{\dot{\tbf{r}}_i} \mathcal{D}(\tbf{r}_h,\dot{\tbf{r}}_h) = \alpha \dot{s} s^{-3} \tbf{q} \nabla_{\tbf{r}_i} U(\tbf{r}_0).
\end{split}
\end{equation*}

In this case, $\alpha$ has dimensions of inverse of time. This relation between 
$\nabla_{\dot{\tbf{r}}_i} \mathcal{D}$ and $\nabla_{\tbf{r}_i} U$ allows us to obtain 
the equation for the homographic parameters ($s(t), \theta(t)$) in the presence of dissipation.

Substituting the homographic form $\tbf{r}_h(t) = \tbf{c}(t) + s(t) \tbf{Q}(\theta(t))(\tbf{r}_0 - \tbf{c}_0)$ 
(Eq. \eqref{eq:homographic}) into the stacked equations of motion (\ref{eq:motion-equation}), and using that $\ddot{\tbf{c}}=0$, 
with identities \eqref{eq:cc-first-dt} and multiplying by $\tbf{Q}^{-1}$, gives
\begin{equation}\label{eq:cc-into-eq-motion}
\begin{split}
    \tbf{M} \lt(\ddot{\tbf{r}}_h - \ddot{\tbf{c}}\rt) 
    &= -\nabla_{\tbf{r}} U(\tbf{r}_h) - \nabla_{\dot{\tbf{r}}} \mathcal{D}(\tbf{r}_h,\dot{\tbf{r}}_h) \\
    \lt(\ddot{s} - s \dot{\theta}^2\rt) \tbf{M} \lt(\tbf{r}_0 - \tbf{c}_0\rt)
    + \lt(2 \dot{s} \dot{\theta} + s \ddot{\theta}\rt) \tbf{J} \tbf{M} \lt(\tbf{r}_0 - \tbf{c}_0\rt)
    &= -\lt(s^{-2} + \alpha \dot{s} s^{-3}\rt) \nabla_{\tbf{r}_0} U(\tbf{r}_0)   \\
\end{split}
\end{equation}
where $\tbf{J} = \textrm{diag}_N(\textbf{j})$.
To satisfy the CC condition, it must hold that 
\begin{equation*}\label{eq:cc-condition}
    \nabla_{\tbf{r}_0} U(\tbf{r}_0) + \lambda \tbf{M} \lt(\tbf{r}_0 - \tbf{c}_0\rt) = 0,
\end{equation*}
noticing that $\lambda = \frac{U(\tbf{r})}{I(\tbf{r})} < 0$.
Simultaneously, since $\tbf{J}$ is a right-angle rotation, it renders 
$\tbf{J}(\tbf{r}_0 - \tbf{c}_0)$ orthogonal to $(\tbf{r}_0 - \tbf{c}_0)$. 
This splits equation (\ref{eq:cc-into-eq-motion}) into terms parallel and orthogonal to 
$(\tbf{r}_0 - \tbf{c}_0)$, yielding a system of equations for the coefficients
\begin{equation}\label{eq:homographic-dissipated}
\begin{cases}
    \ddot{s} - s \dot{\theta}^2 = \lambda \lt(s^{-2} + \alpha \dot{s} s^{-3}\rt) \\
    s \ddot{\theta} + 2 \dot{s} \dot{\theta} = 0
\end{cases}
\end{equation}

Equation (\ref{eq:homographic-dissipated}) is a dissipative Kepler-like system 
in the homographic parameters, with $s(t)$ playing the role of the radial coordinate, 
and $\theta(t)$ the angular position. 
Due to this equivalence, the next section treats the two-body problem under dissipation 
with $d=3$, providing solutions applicable both to the two-body case and to generic 
CCs for arbitrary $N$.

% It is worth noticing that, for whatever other value of $d$, the dissipative force vector in 
% the right hand side of equation (\ref{eq:cc-into-eq-motion}) is not guaranteed to be parallel 
% to the gravitational force. This makes the analysis harder as the the separatarion 
% between terms parallel and orthogonal to $(\tbf{r}_0 - \tbf{c}_0)$ is not clear and 
% it may no longer be valid.

\section{Two bodies}\label{sec:two-bodies}

% region two-bodies-intro
We describe the two-body system in polar coordinates $(r, \theta)$, where $v = \dot{r}$ 
denotes the radial velocity. We choose the mass unit such that the reduced mass is one, 
$m_1 m_2 / (m_1 + m_2) = 1$. The equations of motion take the first-order form
\begin{equation}\label{eq:two-bodies-equations}
\begin{cases}
    \dot{r} = v \\
    \dot{v} = -\partial_r \mathcal{H} - \partial_v \mathcal{D} = \frac{C^2}{r^3} - \gamma \lt(\frac{1}{r^2} + \alpha \frac{v}{r^3}\rt)
\end{cases}
\end{equation}
with $\gamma = G (m_1 + m_2)$. The energy $\mathcal{H}$ and dissipation $\mathcal{D}$ 
functions are given by
\begin{equation}\label{eq:two-bodies-energy-rayleigh}
    \mathcal{H} = \frac{v^2}{2} + \frac{C^2}{2r^2} - \frac{\gamma}{r}
    \qquad\text{and}\qquad
    \mathcal{D} = \frac{\alpha}{2} \frac{\gamma}{r^3} v^2,
\end{equation}
and the (conserved) normalized angular momentum $C$ is defined by
\begin{equation}\label{eq:two-bodies-angular-mom}
    C = r^2 \dot{\theta}.
\end{equation}

It is convenient to define the non-dimensional parameter
\begin{equation}\label{c}
c=\frac{C}{\gamma^{2/3}\alpha^{1/3}}.
\end{equation}
For $\alpha > 0$ and in convenient time and length scales, the dynamics depends 
only on $c$, see equation \eqref{eq:two-bodies-equations-c}.

To understand the solutions of equation \eqref{eq:two-bodies-equations}, we proceed to a 
Poincaré compactification of the phase space in a series of transformations, following \citet{Dumortierbook} 
(Chapter 5.1).
The idea relies on mapping the infinite phase space $(r, v) \in [0, \infty) \times \mathbb{R}$ into a 
sphere, and analyze the dynamics at infinity via stereographic charts (planar maps) covering the sphere.
This allows us to understand the asymptotic behavior of solutions as they approach infinity in the original 
coordinates. 
% endregion

\subsection{Poincaré compactification}

\begin{figure}[H]
    \centering
    \includegraphics[width=0.32\textwidth]{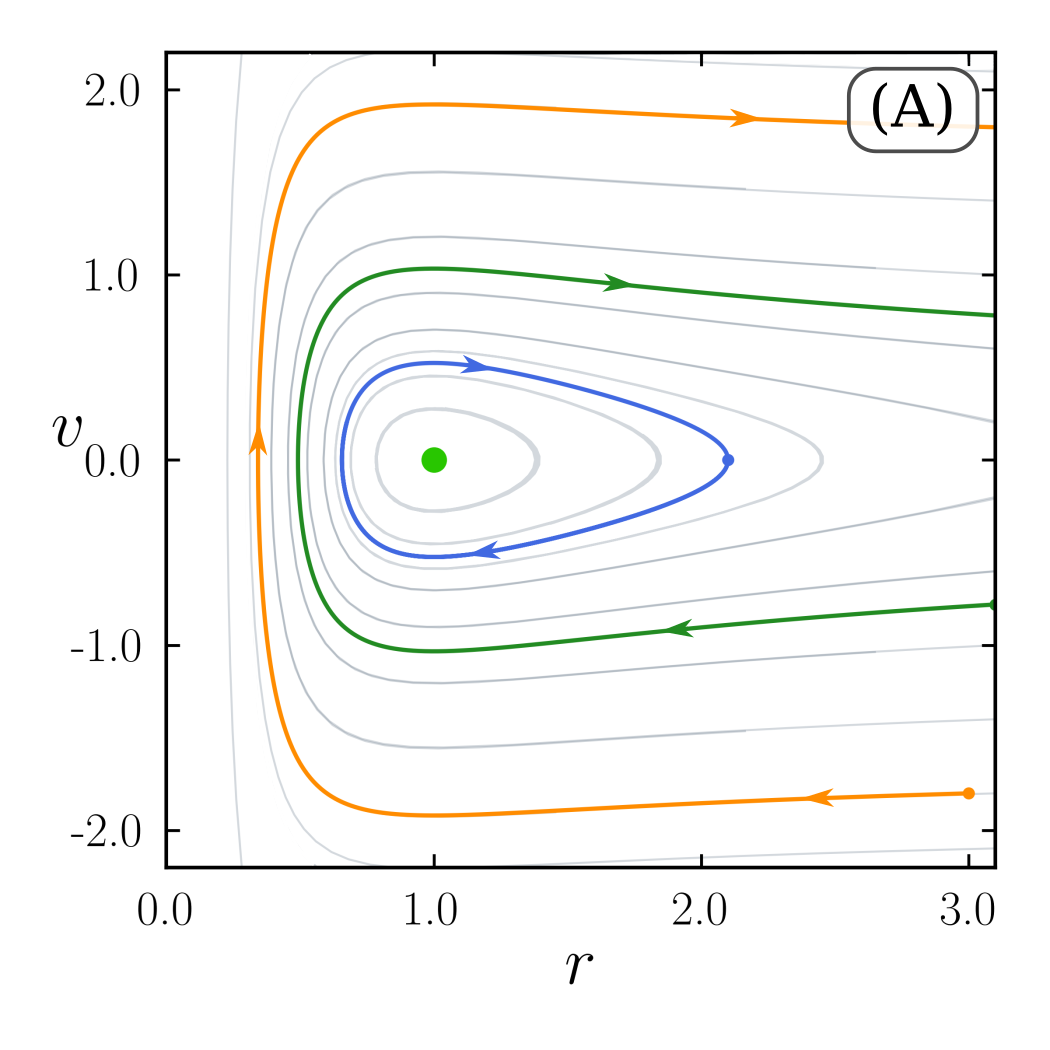}
    \includegraphics[width=0.32\textwidth]{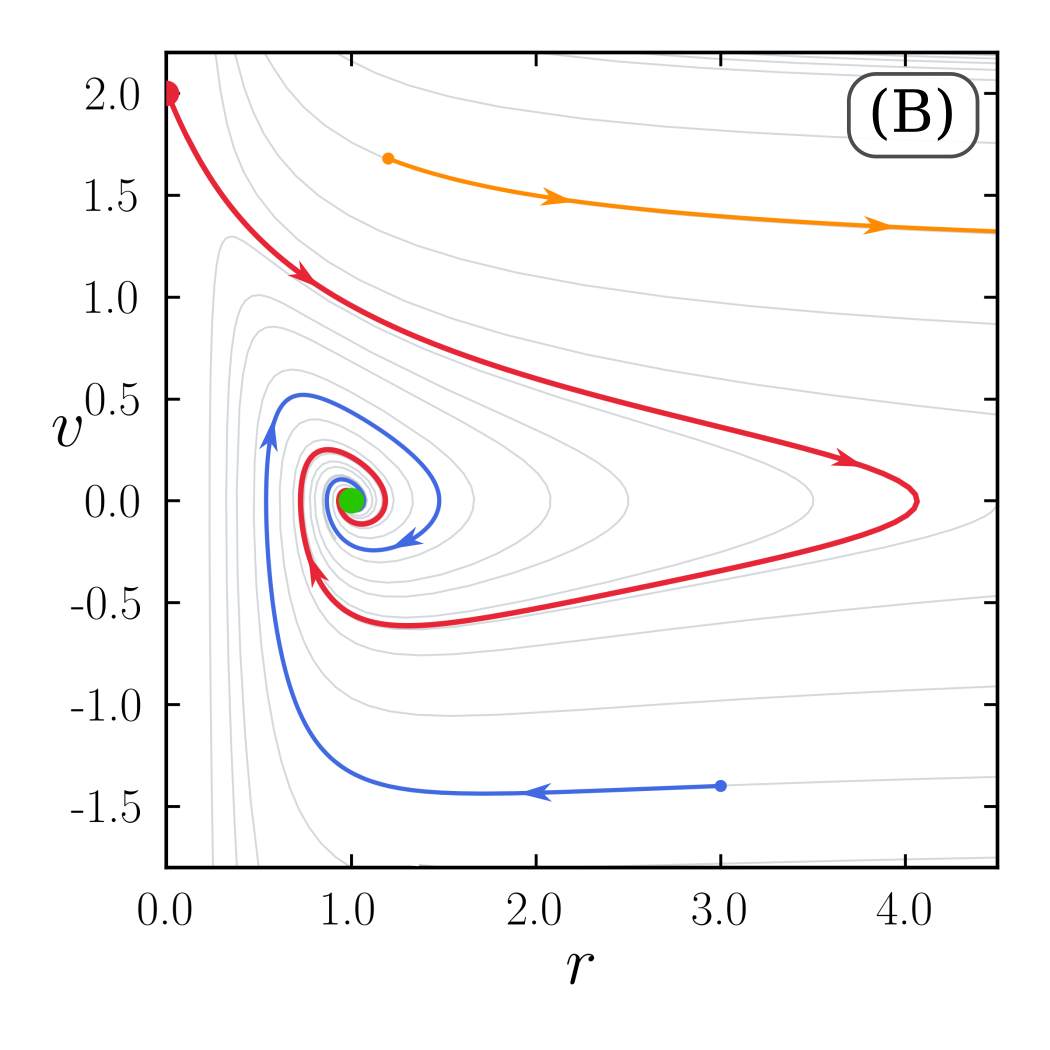}
    \includegraphics[width=0.32\textwidth]{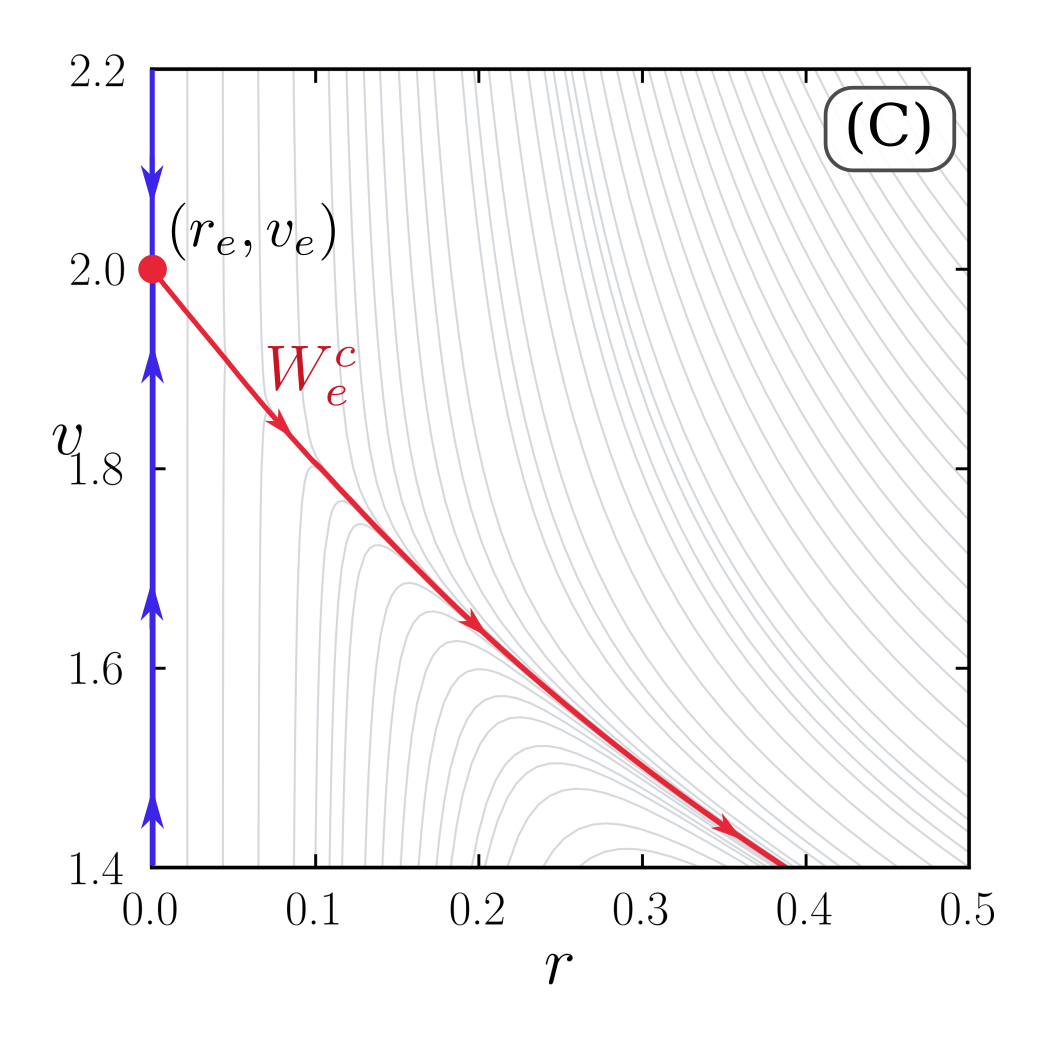}
    \caption{Phase space portraits for the regularized Kepler system (eq. \ref{eq:two-bodies-regularized}) 
             for $C = \gamma = 1$.
            (A) The conservative case ($\alpha = 0$) showing elliptic (blue), 
            parabolic (green), and hyperbolic (orange) orbits.
            (B) The system with dissipation ($\alpha = 0.5$) showing a captured orbit (blue), 
            the weak-unstable-center manifold (red) from the ejection equilibrium $(r_e, v_e) = (0, 2)$, 
            and an escaping orbit (orange); the green dot marks the circular equilibrium 
            $(r_c, v_c) = (1, 0)$. 
            (C) Zoom of (B) near $(r_e, v_e)$; 
            stable branches along $r=0$ are shown in blue.}
    \label{fig:two-bodies-regularized}
\end{figure}

We start by multiplying the equations in (\ref{eq:two-bodies-equations}) by $r^3$ 
to produce a polynomial vector field, which preserves the phase space portrait of the system 
as it is equivalent to a time reparametrization:
\begin{equation}\label{eq:two-bodies-regularized}
\begin{cases}
    \dot{r} = r^3 v \\
    \dot{v} = C^2 - \gamma \lt(r + \alpha v\rt),
\end{cases}
\end{equation}
The vector field (\ref{eq:two-bodies-regularized}) has two equilibria, namely the circular orbit
\begin{equation}\label{eq:circular-orbit}
    (r_c, v_c) = \lt(\frac{C^2}{\gamma}, 0\rt), 
\end{equation}
and an ejection equilibrium
\begin{equation}\label{eq:ejection-orbit}
    (r_e, v_e) = \lt(0, \frac{C^2}{\gamma \alpha}\rt),
\end{equation}
which emerges as an artifact of the $r^3$ scaling and that does not exist 
in the conservative case for $\alpha = 0$ (Fig.~\ref{fig:two-bodies-regularized} (A) and (B)). 

The eigenvalues of the vector field (\ref{eq:two-bodies-regularized}) at the circular orbit $(r_c, v_c)$ are
\begin{equation}\label{eq:eigenvalues-rc}
\begin{split}
    a^c_\pm = -\frac{\alpha\gamma}{2}\lt(1 \pm \sqrt{1 - \frac{4 C^6}{\gamma^4\alpha^2 }}\rt)
            = -\frac{\alpha\gamma}{2}\lt(1 \pm \sqrt{1 - 4 c^6}\rt),
\end{split}
\end{equation}
such that the orbit is an attracting node if $\alpha > \frac{2C^3}{\gamma^2}$ ($c^6<1/4$), 
as $a_{\pm}^c < 0$, and an attracting focus if $0 < \alpha < \frac{2C^3}{\gamma^2}$ ($c^6>1/4$), 
as Im$(a_\pm^c) \neq 0$ produces an oscillatory component (Fig.~\ref{fig:two-bodies-regularized} (B)). 

The dynamics around $(r_e, v_e)$ is less clear. Restricting the vector field (\ref{eq:two-bodies-regularized}) 
over the line $r = r_e = 0$ implies a dynamic
\begin{equation*}
\begin{cases}
    \dot{r} = 0 \\
    \dot{v} = C^2 - \gamma \alpha v
\end{cases}
\end{equation*}
Hence, the equilibrium $(r_e, v_e)$ has an eigenvalue $a^e = -\gamma \alpha$ and a stable 
manifold coinciding with the axis $r = 0$. The other eigenvalue is zero and the associated 
center-manifold $W_e^c$ is tangent to the vector $(\alpha, -1)$ (Fig.~\ref{fig:two-bodies-regularized} (C)).
Locally, 
\begin{equation*}
    W^c_e \approx \left\{v=\frac{C^2}{\gamma \alpha}-\frac{r}{\alpha}\right\},
    \quad \text{for } r > 0 \text{ small}.
\end{equation*}
All orbits with initial conditions $v(0) < 0$ and arbitrarily small $r(0) > 0$ 
must approach $W^c_e$ before entering the circularization spiral, as shown in panel (C).

To proceed and understand the dynamics at infinity we perform a Poincaré compactification of the 
vector field (\ref{eq:two-bodies-equations}) introducing new varibles $(x, y)$ 
with the coordinate transform
\begin{equation*}
    (r, v) \longmapsto (y, x) \quad\text{where}\quad r = \frac{1}{y}, \;\; v = \frac{x}{y},
\end{equation*}
in a way that $y=0$ corresponds to $r = \infty$. This corresponds to the $U_1$ chart 
in \citet{Dumortierbook}, pg. 151.

In the new coordinates, equation (\ref{eq:two-bodies-equations}) becomes
\begin{equation}\label{eq:two-bodies-compact-xy}
\begin{cases}
    \dot{y} = - x y \\
    \dot{x} = C^2 y^4 - x^2 - \gamma (1 + \alpha x) y^3
\end{cases}
\end{equation}
% \textcolor{red}
% {
% (TO OMIT)
% Deriving in time the coordinate change
% \begin{equation*}
% \begin{cases}
%     \dot{r} = -\frac{\dot{y}}{y^2} \\
%     \dot{v} = \frac{\dot{x} y - \dot{y} x}{y^2}
% \end{cases}
% \end{equation*}
% and replacing it back into (\ref{eq:two-bodies-regularized}) 
% \begin{equation*}
% \begin{cases}
%     -\frac{\dot{y}}{y^2} = \frac{1}{y^3} \frac{x}{y} \quad\implies\quad \dot{y} = -\frac{x}{y^2}\\
%     \frac{\dot{x} y - \dot{y} x}{y^2} = C^2 - \gamma \lt(\frac{1}{y} + \alpha \frac{x}{y}\rt)
% \end{cases}
% \end{equation*}
% the second equation develops as
% \begin{equation*}
% \begin{split}
%     \dot{x} y &= -\frac{x^2}{y^2} + C^2 y^2 - \gamma (y + \alpha x y) \\
%     \dot{x} &= C^2 y - \frac{x^2}{y^3} - \gamma (1 + \alpha x) \\
% \end{split}
% \end{equation*}
% Then, multiplying both equations by $y^3$:
% \begin{equation*}
% \begin{cases}
%     \dot{y} = - x y \\
%     \dot{x} = C^2 y^4 - x^2 - \gamma (1 + \alpha x) y^3
% \end{cases}
% \end{equation*}
% }

\begin{figure}[H]
    \centering
    \includegraphics[width=0.32\textwidth]{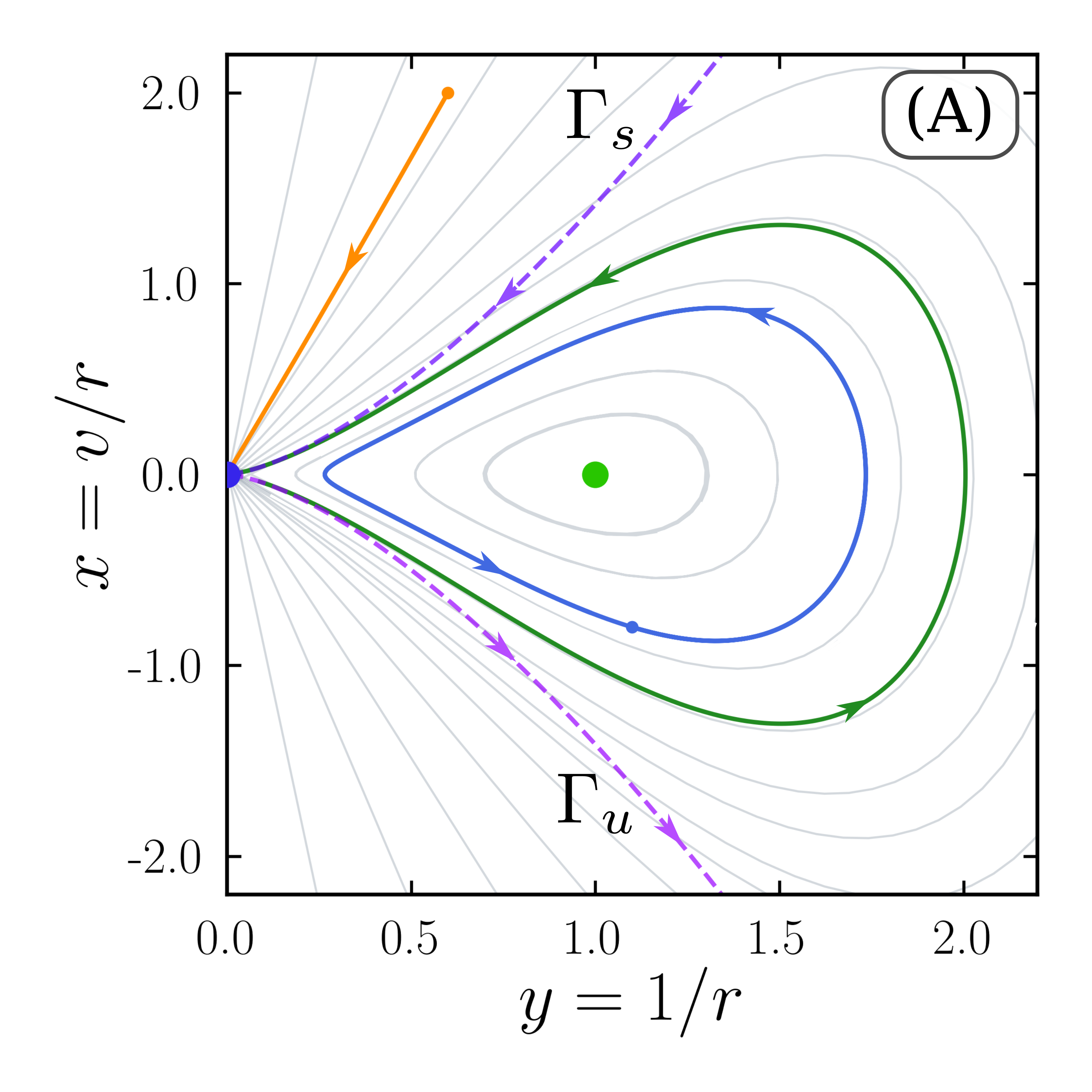}
    \includegraphics[width=0.32\textwidth]{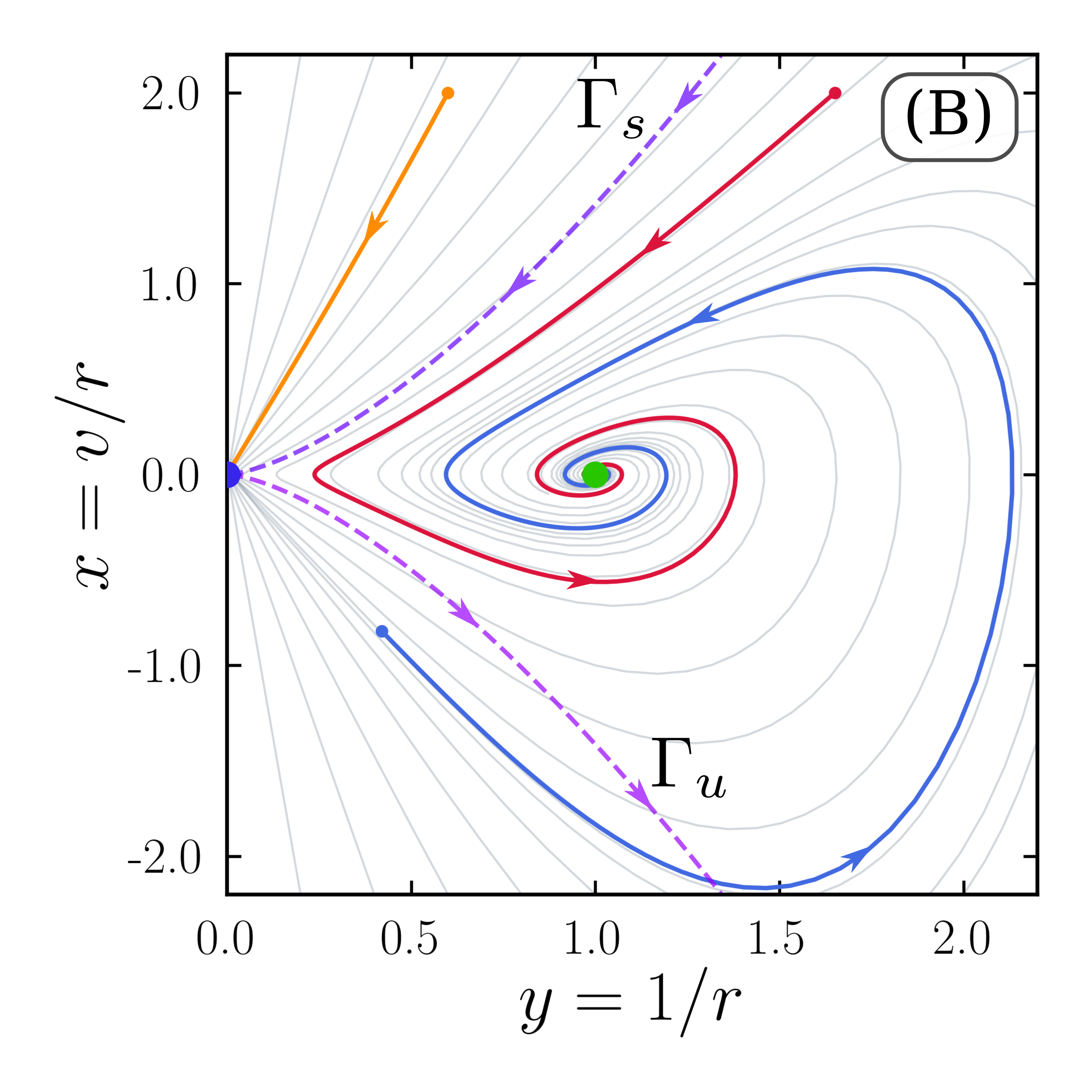}
    \includegraphics[width=0.32\textwidth]{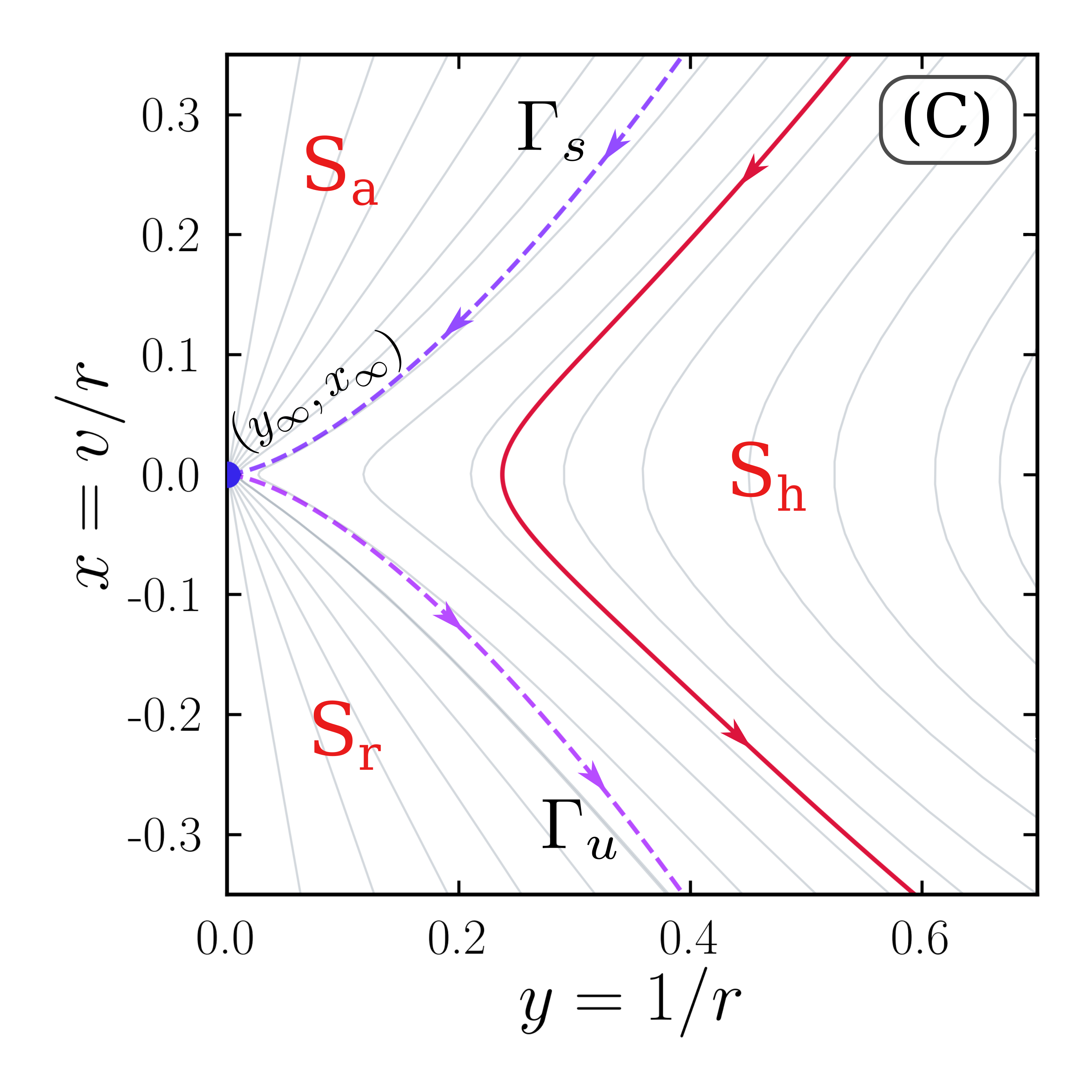}
    \caption{Phase space portraits for the compactified system (\ref{eq:two-bodies-compact-xy}) 
    for $C = \gamma = 1$. 
    (A) The system without dissipation ($\alpha = 0$) showing elliptic (blue), 
    parabolic (green), and hyperbolic (orange) orbits. The green dot is the circular 
    orbit equilibrium $(y_c, x_c) = (1, 0)$.
    (B) The system with dissipation ($\alpha = 0.5$) showing a captured orbit (blue), 
    and an escaping orbit (orange). The curve in red represents the weak-unstable-center 
    manifold from the degenerate ejection equilibrium $(r_e, v_e)$ which here is a point at infinity.
    (C) Zoom of (B) near the degenerate equilibrium $(y_\infty, x_\infty) = (0, 0)$.
    The hyperbolic, attracting, and repelling sectors are denoted $S_h, S_a$, and $S_r$.
    The purple dashed lines $\Gamma_{u(s)}$ are local approximations to the unstable (stable) 
    manifolds of $(y_\infty, x_\infty) = (0, 0)$.}
    \label{fig:two-bodies-compact-xy}
\end{figure}

The new system (\ref{eq:two-bodies-compact-xy}) has two equilibria: 
$(y_c, x_c) = (\frac{\gamma}{C^2}, 0)$, corresponding to the circular orbit, and 
\begin{equation*}\label{eq:compact-equilibrium-infinity}
    r_\infty \coloneqq (y_\infty, x_\infty) = (0, 0).   
\end{equation*}
As the linear approximation of (\ref{eq:two-bodies-compact-xy}) at $(0, 0)$ is 
null, $(y_\infty, x_\infty)$ is a degenerate equilibrium. One may note that all unbounded 
solutions of the Kepler problem ($\alpha = 0$) are homoclinic to $(y_\infty, x_\infty) = (0, 0)$
(see Fig.~\ref{fig:two-bodies-compact-xy} (A)). 

A bifurcation occurs when $\alpha$ exceeds a critical value $\alpha_c$. 
As shown in panel (C) of Fig.~\ref{fig:two-bodies-compact-xy}, for $\alpha>\alpha_c$ 
the weak-unstable-center manifold (red) lies in the hyperbolic sector $S_h$. However, if $\alpha<\alpha_c$
this manifold would lie in the attracting sector $S_a$, for some $\alpha_c \in[0.1,0.5]$. 
At the critical value, the manifold coincides with the stable branch $\Gamma_s$ of the attractor $(y_\infty,x_\infty)$.

To analyze the infinity we blow-up the vector-field at $(y,v)=(0,0)$ using 
\citep[Sec. 3.3]{Dumortierbook}
\begin{equation*}
    (y, x) \longmapsto (p, q) \quad\text{where}\quad y =q^2, \;\; x = pq^3.
\end{equation*}
In the new coordinates the vector field \eqref{eq:two-bodies-compact-xy}, after a 
 multiplication by $2q^{-3}$,  becomes
\begin{equation}\label{eq:two-bodies-compact-qp}
\begin{cases}
    \dot{q} = -q p \\
    \dot{p} =-2 \gamma +2 C^2 q^2-2 \alpha  \gamma p q^3+p^2 
\end{cases}
\end{equation}

The $p$-axis, which corresponds to $q^2=y=\frac{1}{r}=0$, is invariant. There are two hyperbolic 
singularities on this axis: 
\begin{equation*}\label{eq:compact-equilibrium-infinity-qp}
    (q_{\infty +}, p_{\infty +}) = \left(0, \sqrt{2\gamma}\right),\quad\text{and}\quad
    (q_{\infty -}, p_{\infty -}) = \left(0, -\sqrt{2\gamma}\right),
\end{equation*}
connected by a solution from $(q_{\infty +}, p_{\infty +})$ to $(q_{\infty -}, p_{\infty -})$. The whole 
axis $p$ comes from the blow-up of the equilibrium $(y_\infty, x_\infty)$.

The equilibrium $(q_{\infty +}, p_{\infty +})$ has a stable manifold $\Gamma_s$
that, up to linear approximation, is perpendicular to the $p$-axis: $\Gamma_s\approx 
\{q,p=\sqrt{2\gamma}\}$ for $|q|$ small.
Similarly, $(q_{\infty -}, p_{\infty -})$ has an unstable manifold $\Gamma_u$
that, up to linear approximation, is perpendicular to the $p$-axis:
$\Gamma_u\approx \{q,p=-\sqrt{2\gamma}\}$ for $|q|$ small.

In the original variables ($y, x$), we then have a local expression for the manifold branches
\begin{equation}\label{eq:xy-manifolds}
       \Gamma_{s} = \sqrt{2\gamma} \, y^{\frac{3}{2}} \,, \qquad 
       \Gamma_{u} = -\sqrt{2\gamma} \, y^{\frac{3}{2}} \,, \qquad y>0, 
\end{equation}
\\\noindent
which does not depend on $C$ or $\alpha$.
Later in this analysis we will show that these manifolds are related to circularization and escape basins of attraction.

For $\alpha = 0$, the vector field (\ref{eq:two-bodies-compact-xy}) has the first integral 
$\mathcal{H} = \frac{y^3 \lt(C^2 y - 2\gamma \rt) + x^2}{2 y^2}$ (from eq. \ref{eq:two-bodies-energy-rayleigh}), 
and both characteristic curves $\Gamma_u$ and $\Gamma_s$ are determined by the equation $\mathcal{H} = 0$.
One can check that this leads to the approximation in equation (\ref{eq:xy-manifolds}).

In Fig.~\ref{fig:two-bodies-compact-xy} we show the local structure of the orbits 
of the vector field \eqref{eq:two-bodies-compact-xy} for some  values of $\alpha$.
The stable curve
$\Gamma_s$ approaching  $(y_\infty, x_\infty)$ for positive time, and 
the curve $\Gamma_u$ approaching it for 
negative time divide the phase space $x$--$y$ into three sectors: a hyperbolic sector $S_h$, 
an attracting sector $S_a$, and a repelling sector $S_r$ 
(see Fig.~\ref{fig:two-bodies-compact-xy} (C)).

In the Kepler problem, orbits with positive energy are asymptotically free, with $\mathcal{H} \to \frac{v^2}{2}$ 
as $t \to \pm\infty$. We define {\sl scattering orbits}, in the dissipative or conservative scenarios, as those for 
which the following two limits exist
\begin{equation*}
    \lim_{t\to-\infty} v(t) = -v_\psi < 0 \quad\text{and}\quad \lim_{t\to\infty} v(t) = v_\omega > 0.
\end{equation*}

In order to compute the asymptotic shape of a scattering orbit close to a point at infinity,
one can use that
\[\lim_{t\to\pm\infty} \mathcal{D} = \lim_{t\to\pm\infty} \frac{\alpha}{2} \frac{\gamma}{r^3} v^2 = 0\]
to neglect dissipation of energy and solve
\begin{equation*}
    \mathcal{H} = \frac{y^3 \lt(C^2 y - 2\gamma \rt) + x^2}{2 y^2} = \frac{v^2}{2}, \quad\text{or}\quad
    y^3 \lt(C^2 y - 2\gamma \rt) + x^2 = v^2 y^2,
\end{equation*}
near $(y, x) = (y_\infty, x_\infty) = (0, 0)$. Up to leading order, the solution is
\begin{equation*}
    x = \pm v y + \mathcal{O}(y^{3/2}).
\end{equation*}

Therefore, a scattering orbit leaves the point at infinity tangent to the line $x = -v_\psi y$ 
through the repelling sector $S_r$, and arrives at the same point tangent to the line $x = v_\omega y$ 
through the attracting sector $S_a$ (see Fig.~\ref{fig:two-bodies-compact-xy} (C)).
In the presence of dissipation, $|v_\omega| < |v_\psi|$.

% Besides, the equation for the characteristic curves $\Gamma_{u(s)}$, $x = \pm \sqrt{2\gamma} y^{\frac{3}{2}}$ implies 
% that scattering orbits arrive and leave $(x_\infty, y_\infty)$ tangent to the $y$--axis. Therefore, as expected, 
% scaterring orbits must both leave the point at infinity by the repelling sector $S_r$ and arrive at the point 
% at infinity in the attracting sector $S_a$ (see Fig.~\ref{fig:two-bodies-compact-xy} (C)).

To understand the dynamics at the remaning part of infinity, we will make use of a second chart 
to cover the points $(r, v) = (0, \pm \infty)$ (the $U_2$ chart in \citet{Dumortierbook}, pg. 152).
The change of variables from (\ref{eq:two-bodies-equations}) in this case is
\begin{equation}\label{rv}
    (r, v) \mapsto (w, z) \quad\text{where}\quad w = \frac{r}{v}, \quad z = \frac{1}{v}
\end{equation}
In these coordinates, $z = 0$ corresponds to $v = \pm\infty$. At first we 
will be interested on the limit as $v\to\infty$ that corresponds to  the quadrant
$w>0$, $z>0$.

After multiplying by $w^3$, the vector field (\ref{eq:two-bodies-equations}) in the new chart becomes
% TO OMIT
% \begin{equation*}
% \begin{split}
%     \begin{cases}
%     \dot{r} = r^3 v \\
%     \dot{v} = C^2 - \gamma \lt(r + \alpha v\rt)
%     \end{cases} \\
%     \begin{cases}
%     \dot{w} &=  \frac{\dot{r} v - r \dot{v}}{v^2} \\
%     \dot{z} &= -\frac{\dot{v}}{v^2}
%     \end{cases} \\
%     \begin{cases}
%     \dot{w} &= \frac{r^3 v^2 - r C^2 + \gamma r \lt(r + \alpha v\rt)}{v^2} \\
%     \dot{z} &= -\frac{C^2 - \gamma \lt(r + \alpha v\rt)}{v^2}
%     \end{cases} \\
%     \begin{cases}
%     \dot{w} &=  \frac{w^3 z^{-3} z^{-2} - w z^{-1} C^2 + \gamma w z^{-1} \lt(w z^{-1} + \alpha z^{-1}\rt)}{z^{-2}} \\
%     \dot{z} &= -\frac{C^2 - \gamma \lt(w z^{-1} + \alpha z^{-1}\rt)}{z^{-2}}
%     \end{cases} \\
%     \begin{cases}
%     \dot{w} &=  w^3 z^{-3} - \lt(C^2 w z - \gamma w \lt(w + \alpha\rt)\rt)  \\
%     \dot{z} &= -z\lt(C^2 z - \gamma \lt(w + \alpha\rt)\rt)
%     \end{cases} \\
%     \begin{cases}
%     \dot{w} &=  w^3 z^{-3} - w \lt(C^2 z - \gamma \lt(w + \alpha\rt)\rt)  \\
%     \dot{z} &= z\lt(\gamma \lt(w + \alpha\rt) - C^2 z\rt)
%     \end{cases} \\
%     \begin{cases}
%     (\times z^3) \quad \dot{w} &= w\lt(w^2 + z^3\lt(\gamma \lt(w + \alpha\rt) - C^2 z\rt)\rt)  \\
%     (\times z^3) \quad \dot{z} &= z^4\lt(\gamma \lt(w + \alpha\rt) - C^2 z\rt)
%     \end{cases} \\
% \end{split}
% \end{equation*}
\begin{equation}\label{eq:two-bodies-compact-wz}
\begin{cases}
    \dot{w} = w\lt(w^2 + z^3\lt(\gamma \lt(w + \alpha\rt) - C^2 z\rt)\rt)  \\
    \dot{z} = z^4\lt(\gamma \lt(w + \alpha\rt) - C^2 z\rt)
\end{cases}
\end{equation}
which has two equilibria: $(w_e, z_e) = (0, \frac{\alpha \gamma}{C^2})$, which corresponds to 
the $(r_e, v_e)$ equilibrium previously found, and a new equilibrium
\begin{equation}\label{vinf}
    v_\infty:= (w_{v\infty}, z_{v\infty}) = (0, 0).
\end{equation}
The circular equilibrium $(r_c, v_c)$ is mapped to infinity in this chart.

\begin{figure}[H]
    \centering
    \includegraphics[width=0.48\textwidth]{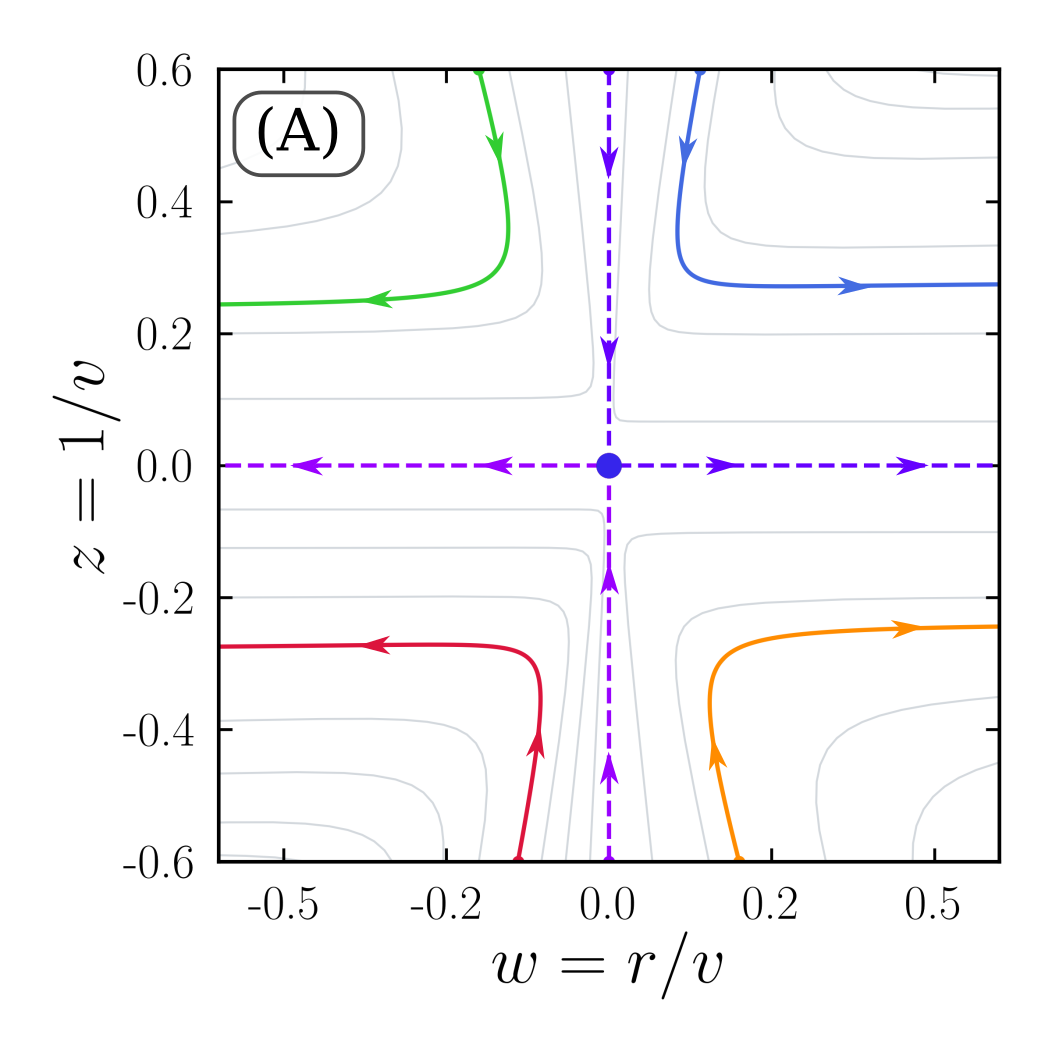}
    \includegraphics[width=0.48\textwidth]{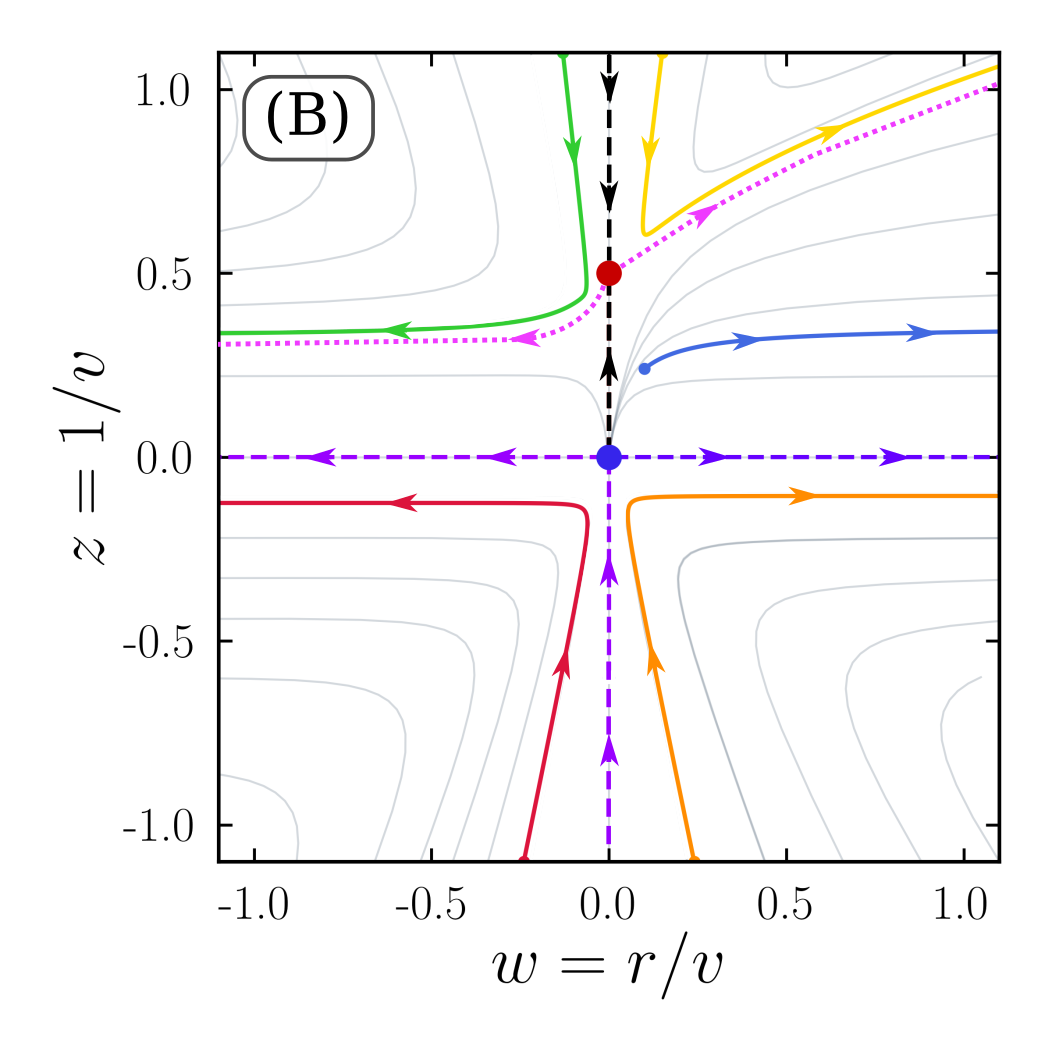}
    \caption{Phase space portraits for the compactified system (\ref{eq:two-bodies-compact-wz}) for $C = \gamma = 1$.
    (A) The system without dissipation ($\alpha = 0$). Dashed lines show the stable and unstable manifolds 
    for $(w_{v\infty}, z_{v\infty}) = (0,0)$ (blue point) and solid lines a few illustrative orbits.
    (B) The system with dissipation ($\alpha = 0.5$). The equilibrium $(w_e, z_e) = (0, 0.5)$ (red point)
    divides the phase space, as indicated by its manifolds (dotted pink lines). }
    \label{fig:two-bodies-compact-wz}
\end{figure}

In Fig.~\ref{fig:two-bodies-compact-wz} we show phase portraits for the system with and without 
dissipation. The correctness of this numerically computed phase portrait is corroborated 
by the following analysis.

We blow up the vector field at $ (w_{v\infty}, z_{v\infty}) = (0, 0)$ using 
\citep[Sec. 3.3]{Dumortierbook}
\begin{equation}\label{pqv}
    (w, z) \longmapsto (p, q) \quad\text{where}\quad w = pq^3, \;\; z =q^2,
\end{equation}
to obtain, after division by $q^6/2$,
\begin{equation}\label{eq:two-bodies-compact-qp-2}
\begin{cases}
    \dot{p} =  p  \left(2 p^2 - \alpha \gamma + C^2 q^2 - \gamma p q^3\right) \\
    \dot{q} = q \left(\alpha  \gamma -C^2 q^2+\gamma  p q^3\right)\\
\end{cases}
\end{equation}
In the region $p\ge 0$, $q\ge 0$, this equation has three equilibria: $(p,q)=(0,0)$,  $(p_e,q_e)=(0,\sqrt{\frac{\alpha \gamma}{C^2}})$, which corresponds to the ejection point, 
and a new one $(p,q)=(\sqrt{\frac{\alpha \gamma}{2}},0)$.

Both equilibria $(p,q)=(0,0)$ and $(p,q)=\lt(\sqrt{\frac{\alpha \gamma}{2}},0\rt)$ are hyperbolic.
The equilibrium $(p,q)=(0,0)$ is a saddle, with its unstable manifold lying on the $q$-axis and its 
stable manifold along the $p$-axis. The equilibrium $(p,q)=\lt(\sqrt{\frac{\alpha \gamma}{2}},0\rt)$ is 
a source with strong unstable manifold on the $p$-axis (eigenvalue $2\alpha \gamma$)
and weak unstable manifold locally given by 
$W^{ws}_\infty\approx  \{p=\sqrt{\frac{\alpha \gamma}{2}}, q\}$ for $|q|$ small (eigenvalue $\alpha \gamma$).

In the variables $(w,z)$, the positive branch of $W^{ws}_\infty$ is given by 
\[
W^{ws}_\infty\approx \left\{w=\sqrt{\frac{\alpha \gamma}{2}} z^{3/2}\right\}\,,\qquad z > 0 \ \ \text{small},
\]
while in the original $(r,v)$ variables it corresponds to
\[
W^{ws}_\infty\approx \left\{v=\frac{\alpha \gamma}{2}\frac{1}{r^2}\right\} \,,\qquad r > 0 \ \ \text{small}.
\]

Finally, we analyze the limit $v \to -\infty$ that,  after 
the change of variables in equation \eqref{rv}, corresponds to 
the  quadrant $w\le 0$, $z\le 0$ in equation \eqref{eq:two-bodies-compact-wz}.
We denote the  equilibrium at $v=-\infty$ as
\begin{equation*}
    (w_{v-\infty}, z_{v-\infty}) = (0, 0)
\end{equation*}

A desingularization of $ (w_{v-\infty}, z_{v-\infty})$ is achieved 
using variables similar to those in equation \eqref{pqv} 
\begin{equation}
    (w, z) \longmapsto (p, q) \quad\text{where}\quad w = pq^3, \;\; z =-q^2
\end{equation}
We are interested in the quadrant $p\le 0$, $q\ge 0$.
 
In the new variables, the vector field in equation \eqref{eq:two-bodies-compact-wz}
becomes, after division by $q^6/2$,
\begin{equation}\label{eq:two-bodies-compact-qp-3}
\begin{cases}
    \dot{p} =  p  \left(2 p^2+\alpha  \gamma +C^2 q^2+\gamma  p q^3\right) \\
    \dot{q} = -q \left(\alpha  \gamma +C^2 q^2+\gamma  p q^3\right)\\
\end{cases}
\end{equation}

This equation has a single equilibrium $(p,q)=(0,0)$, which is a saddle.
Its unstable manifold lies on the $p$-axis, and its 
stable manifold lies on the $q$-axis. As expected, the outcome of this analysis is the phase space 
portrait shown in the quadrant $w\le 0$, $z\le 0$ in Fig.~\ref{fig:two-bodies-compact-wz} \footnote{An information provided by the blow up is that in the quadrant 
$w\le 0$, $z\le 0$ solutions initially close to the equilibrium $(0,0)$ approach it 
approximately along straight lines 
$z(t)\approx \frac{z(0)}{w(0)} w(t)$ before diverging along the $w$-axis. This can be observed in 
Fig.~\ref{fig:two-bodies-compact-wz}.}.

\subsubsection{Phase portrait topology}

To summarize, we recall that the initial vector field (\ref{eq:two-bodies-equations}) describing the radial dynamics of 
the Kepler problem was regularized via a time rescaling (multiplication by $r^3$) yielding vector field (\ref{eq:two-bodies-regularized}).
Since this vector field is polynomial, it admits a Poincaré compactification onto the disk, 
where only the half-disk with $r > 0$ is relevant to the analysis of the dissipative Kepler problem.
The phase portrait topology of this half-disk was assembled from three different charts, each 
one obtained via local projections by compactification transforms.

Inspired by \cite{Diacu1999}, solutions of \eqref{eq:two-bodies-equations} can be placed in 
the categories defined as: 
(I) {\sl capture} orbit if $r\to\infty$ in the past and $r$ is bounded in the future, 
(II) {\sl escape} orbit if $r\to\infty$ in the future and $r$ is bounded in the past, 
(III) {\sl scattering} orbit if $r\to\infty$ in the past and the future,
(IV) {\sl ejection} orbit if $r\to 0$ in the past and $r$ bounded in the future, 
(V) {\sl bounded} orbit if $r$ is bounded in the past and in the future, and
(VI) {\sl collision} orbit if $r\to 0$ in the future, to be described in section \ref{Sec:Coll}.

\begin{figure}[H]
    \centering
    \includegraphics[width=0.32\textwidth]{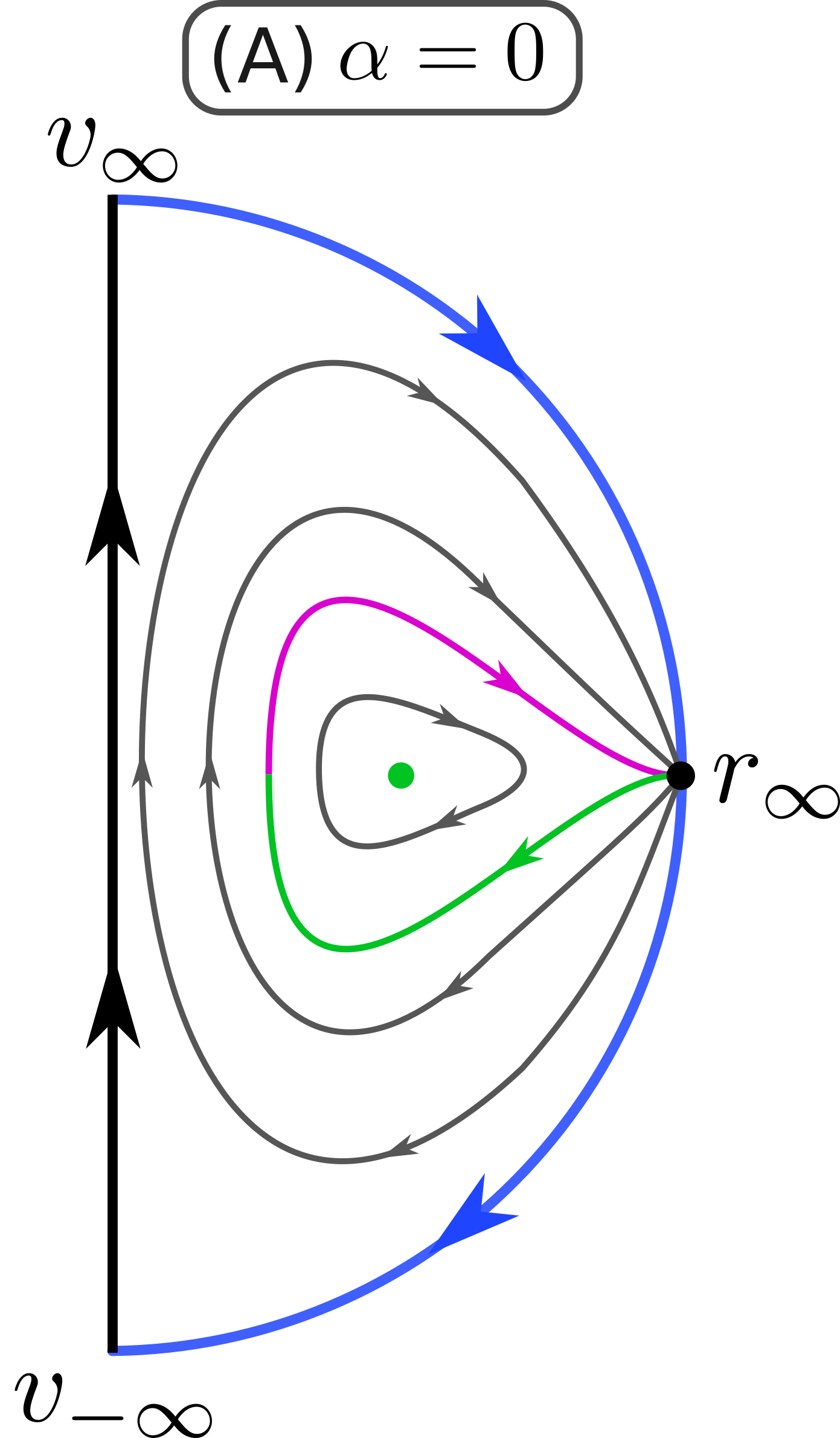}
    \includegraphics[width=0.32\textwidth]{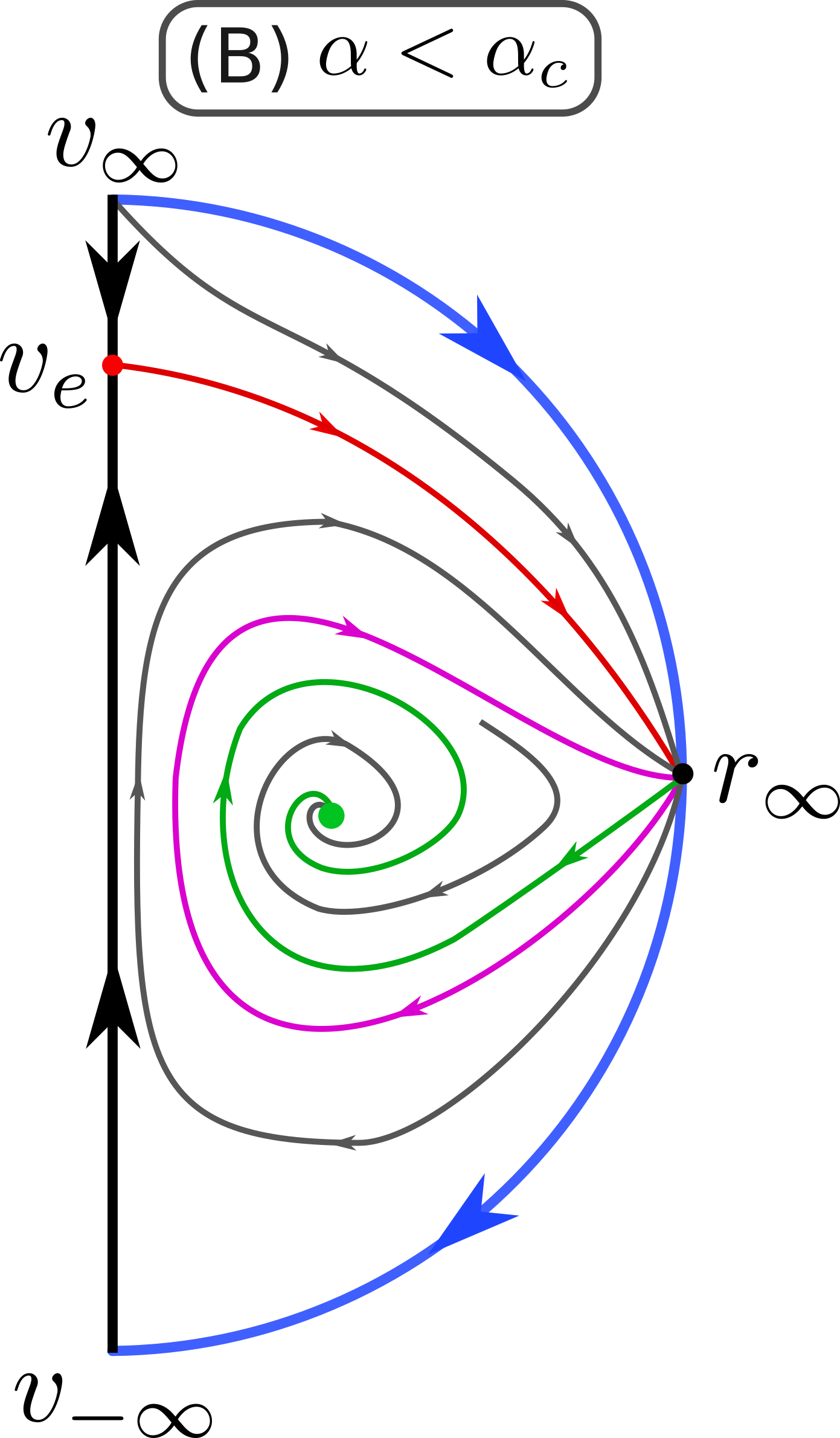}\\\vspace{0.2cm}
    \includegraphics[width=0.32\textwidth]{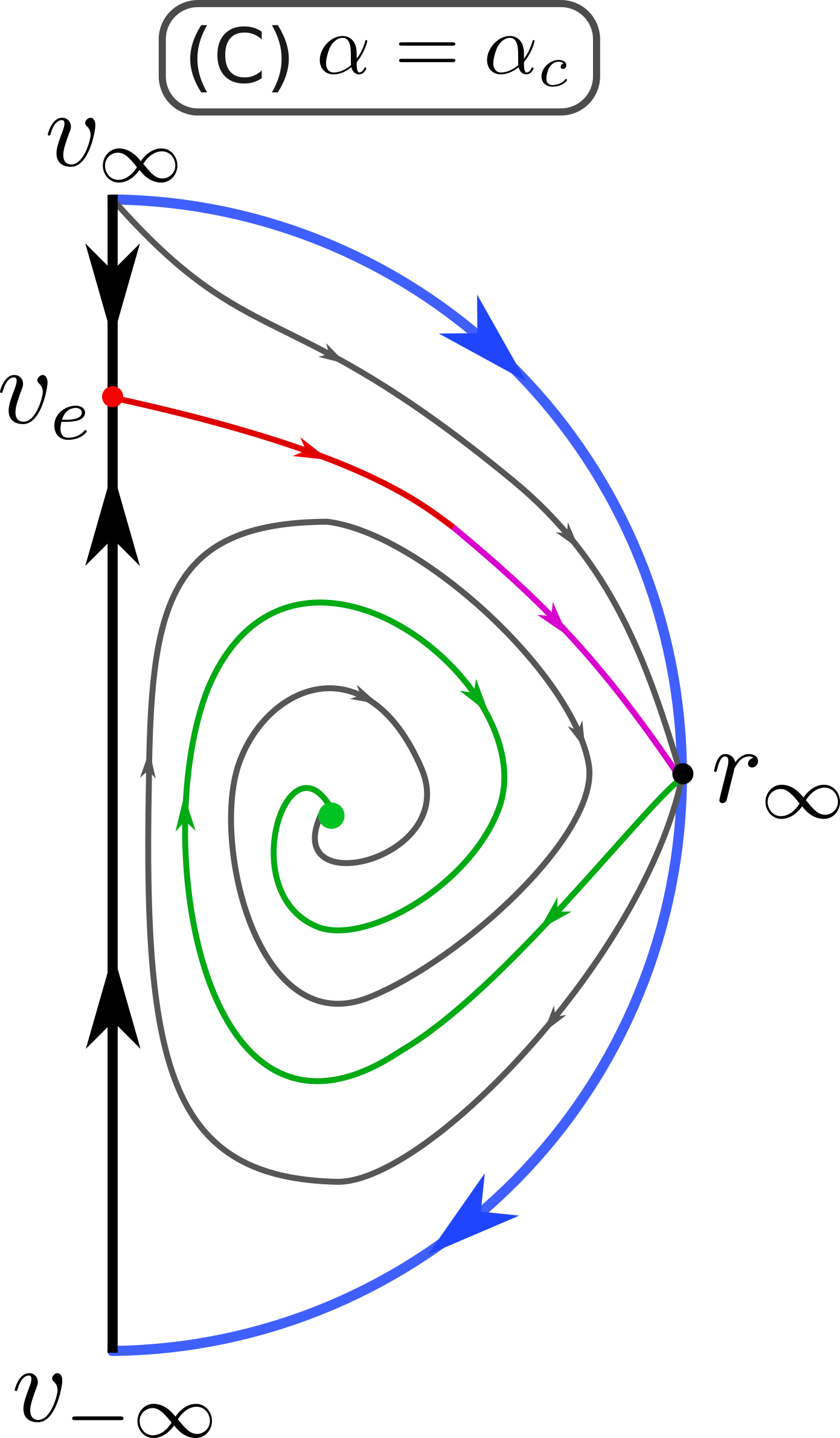}
    \includegraphics[width=0.32\textwidth]{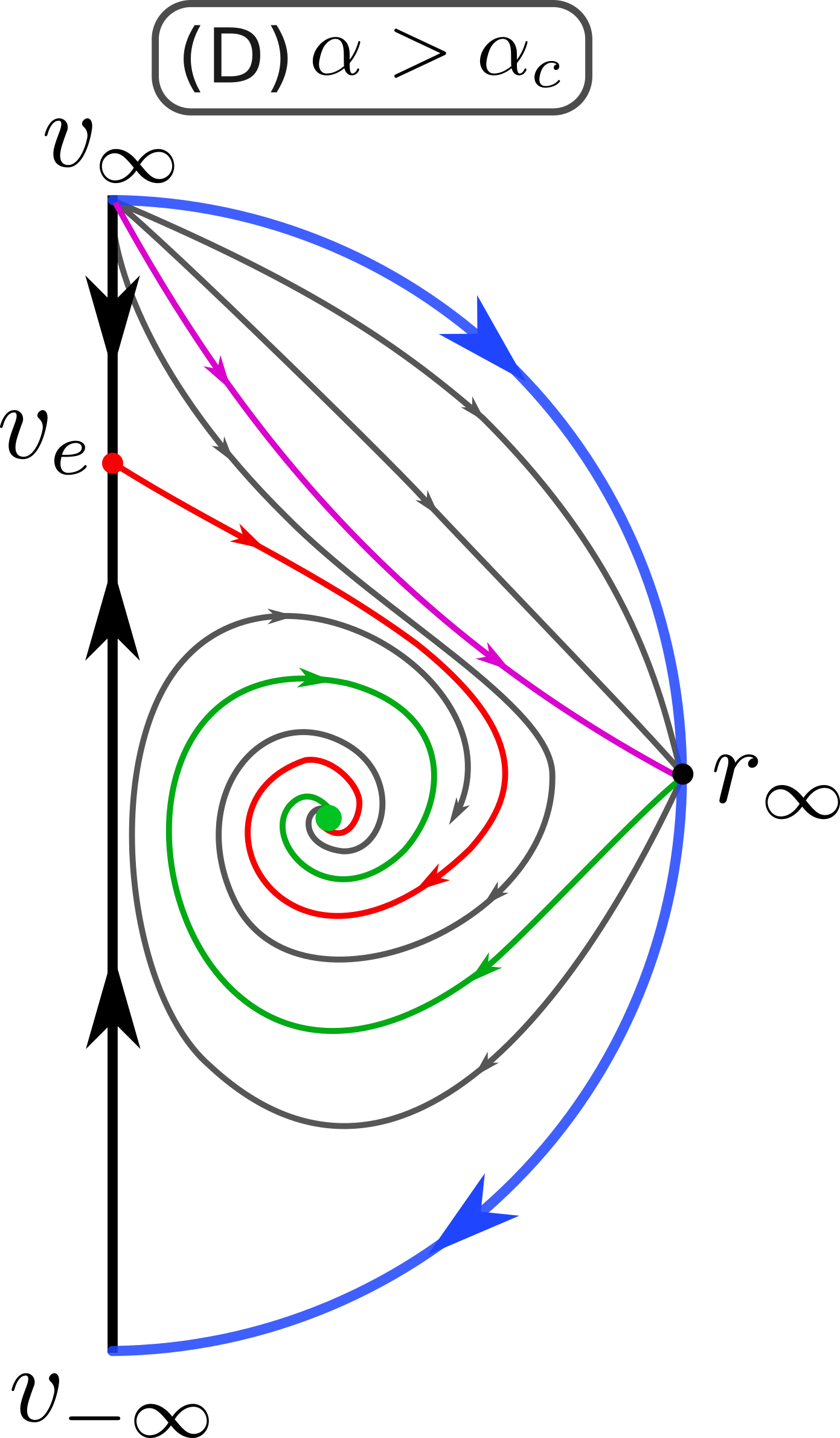}
    \caption{Sketches of the flow line topology on the Poincaré disk for the Kepler 
             system. Green dots mark the circular equilibrium $(r_c, v_c)$. 
             (A) Conservative case ($\alpha = 0$ or $c = \infty$), showing the parabolic separatrix, 
             which corresponds to the merging of the stable $\Gamma_s$ (pink) and unstable $\Gamma_u$ (green) curves of the 
             ``equilibrium'' $r_\infty$, see equation \eqref{eq:compact-equilibrium-infinity}. 
             Orbits are either bounded or scattered.
             (B) Weakly dissipative case ($\alpha = 0.1$ or $c\approx 2.15$), showing that $\Gamma_s$ (pink) is homoclinic to $r_\infty$, 
             while $\Gamma_u$ (green) is attracted to $(r_c, v_c)$. 
             All points inside $\Gamma_s$ are capture orbits, i.e. they converge to the circular orbit as $t\to\infty$ and diverge to $r=\infty$ as $t\to -\infty$. 
             The center-unstable manifold $W^c_e$ (red) of the ejection point $v_e$ converges to $r_\infty$ as $t\to\infty$. 
             Orbits outside $\Gamma_s$ and below $W^c_e$ belong to the scattering region, being bi-asymptotic to $r_\infty$ as $t\to\pm\infty$. 
             Orbits above $W^c_e$ form the escaping region, being asymptotic to $r_\infty$ as $t\to\infty$ and to $v_\infty$, as $t\to -\infty$, see equation \eqref{vinf}. 
             (C) Critic dissipation case ($\alpha = \alpha_c \in (0.1, 0.5)$) at which $W^c_e$ merges with $\Gamma_s$; at this bifurcation, the set 
             of scattering orbits (present for $0\le \alpha<\alpha_c$) disappears.
             (D) Strongly dissipative case ($\alpha = 0.5$ or $c\approx 1.26$), showing that $\Gamma_s$ (pink) is asymptotic to $v_\infty$ as $t\to -\infty$. 
              Orbits above $\Gamma_s$ belong to the escaping region, while those below $\Gamma_s$ lie in the basin of attraction of the circular orbit. 
              The center-unstable manifold $W^c_e$ (red) is attracted to the circular orbit as well as $\Gamma_u$ (green). 
              Orbits below $\Gamma_s$ and ``above'' $W^c_e\cup \Gamma_u $ are asymptotic to $v_\infty$ as $t\to -\infty$. 
              Orbits  ``below'' $W^c_e \cup \Gamma_u $ are asymptotic to $r_\infty$ as $t\to -\infty$.}
    \label{fig:topologies}
\end{figure}

To build the full phase portrait topology on the Poincaré disk, we considered 
four scenarios: $\alpha = 0$, $\alpha = 0.1 < \alpha_c$, $\alpha = \alpha_c$, and $\alpha = 0.5 > \alpha_c$, where $\alpha_c$ is 
the critical dissipation at bifurcation, all for $C = \gamma = 1$ (see Fig.~\ref{fig:topologies}).
In terms of the non-dimensional parameter of the system $c=\frac{C}{\gamma^{2/3}\alpha^{1/3}}$, equation
\eqref{c}, these parameters correspond respectively to: $c=\infty$, $c\approx 2.15$, and 
$c\approx 1.26$.
We remark that the basin of attraction of $r_c$ increases along with $\alpha$, as expected.
% The same topological results may be obtained for any other values of $C, \gamma > 0$.

Other radial exponents $d > 1$ may result in the same topology for the dynamics as $d = 3$, 
requiring only appropriate time rescaling when the vector field is made polynomial for compactification (see Eq.~\eqref{eq:two-bodies-regularized}).
However, the correspondence to central configurations may be lost as the spatial symmetry between dissipation and 
gravity forces is broken.
In general, it is unclear whether the lack of this symmetry yields a dynamics on the same 
invariant set as the one preserved when $d=3$. For $d \neq 3$, the most plausible candidates would be 
equal-mass or regular $n$-gon configurations.

\cite{Margheri2020} analyzed a non-central drag with singular distance dependence over a 
wide range of exponents; although their model does not conserve angular momentum and is 
not directly comparable to ours, their methods may give a hint on how to adapt the Kepler system 
for any $d > 0$.

An analysis of the dynamics near the ejection equilibrium $(r_e, v_e)$ and at infinity was made by \cite{Diacu1999} for 
a generic Stokes' drag, closely comprising our model as a particular case (when $\beta = 0$ in their notation). 
However, as the purely radial case was not treated, his results are qualitatively different from ours, as circularization is not 
possible in his case, with orbits either escaping or colliding (even for $C \neq 0$).

\subsection{The angular-momentum-energy space}

\begin{figure}[H]
    \centering  
    \includegraphics[width=0.6\textwidth]{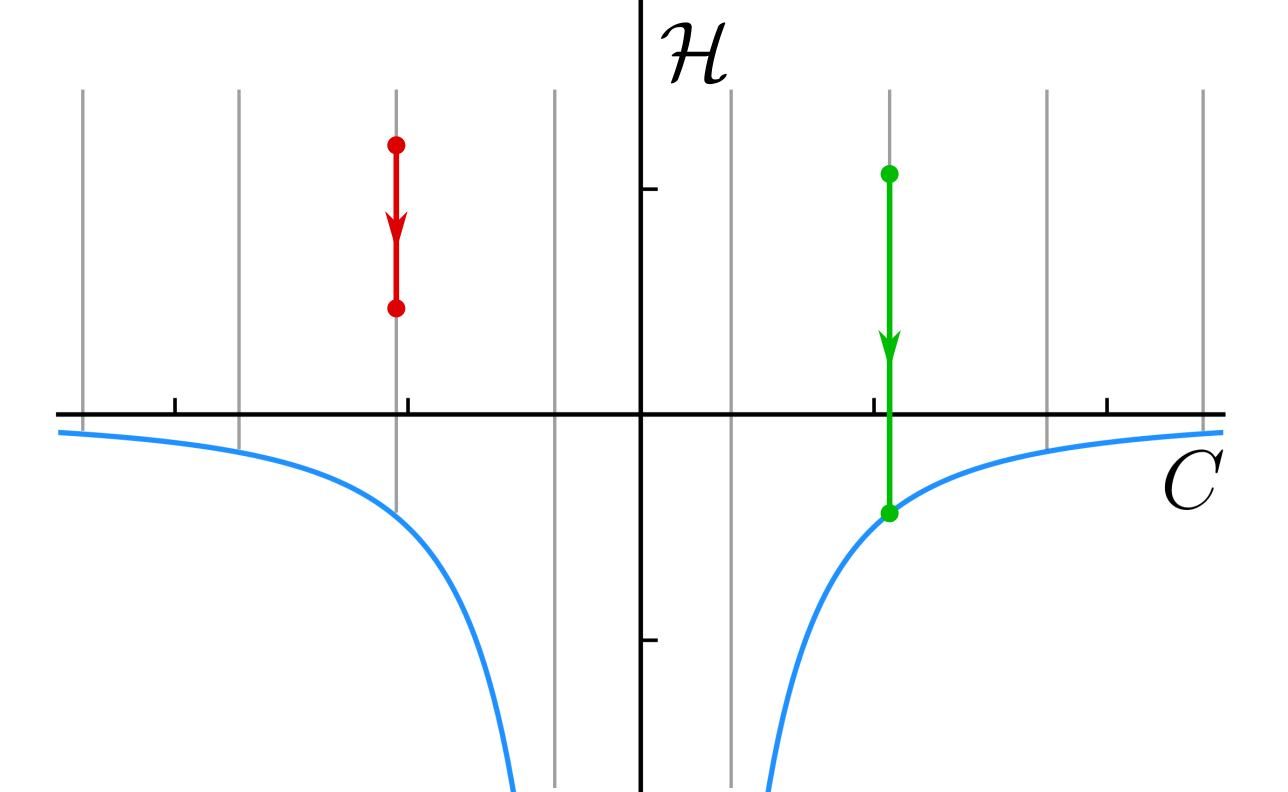}
    \caption{Curve of minimum energy $\mathcal{H}$ as a function of the angular momentum $C$, 
             achieved at the circular orbit. 
             Every orbit is mapped as a segment on the vertical lines above this curve. 
             In green, an orbit captured by the circular attractor. 
             In red, an example of a scattered (escape) orbit.}
    \label{fig:CH-map}
\end{figure}

The angular-momentum-energy map, or CH map, is defined as the map that asso\-ciates to each state 
$(\tbf{r}, \tbf{v}) \in \mathbb{R}^2 /\{0\} \times \mathbb{R}^2$ of the dissipative Kepler problem 
the pair $(C = (\tbf{r}\times\tbf{v})\cdot\hat{\tbf{z}}, \mathcal{H}) \in \mathbb{R}^2$.
For a given $C$, the minimum of $\mathcal{H} = \frac{v^2}{2} + \frac{C^2}{2 r^2} - \frac{\gamma}{r}$ is achieved 
at the circular orbit, $r_c = \frac{C^2}{\gamma}$ and $v = 0$, with energy $\mathcal{H}_\text{min} = -\frac{\gamma^2}{2 C^2}$.
Thus, the image of the CH map is the region above the graph in Fig.~\ref{fig:CH-map}.

An orbit in the conservative Kepler system maps to a single point in the CH space, as $C$ and $\mathcal{H}$ are constants. 
In the presence of dissipation, however, the image of an initial point, not on the curve of 
minimum energy, always moves vertically down.
There are two possibilities for this motion as $t \to \infty$. One, where the point moves vertically down till it 
reaches the minimum energy curve, which happens if, and only if, the initial point is in the basis of attraction of the 
circular orbit.
The second possibility is that the point moves vertically down until reaching a limit with $\mathcal{H} > 0$.
This scenario happens if, and only if, the initial point belongs either to the scattering region or to the escaping region.

\subsection{Collisions and almost collisions.}
\label{Sec:Coll}

In the dissipative Kepler problem considered here, collisions differ significantly from those in the conservative case. 

Collisions occur only when the angular momentum $C$ is zero. In this case, independently of the value of the dissipative parameter $\alpha\ge 0$, the orbits lie on straight segments with the origin, which is the collision point, as an endpoint.

In the conservative case, an almost collision corresponds to an elliptic orbit with eccentricity $e$ close to one and one focus at the origin. Since $C=\sqrt{\gamma a(1-e^2)}$, where $a$ is the semimajor axis, we have $C\approx 0$. The orbit then approximates a segment of length $2a$, with one endpoint (the periapsis) close to the origin. This suggests a collision regularization procedure: for $C=0$, the orbit becomes a segment of length $2a$ with the origin as one endpoint, and the particle moves periodically between the endpoints with infinite velocity at the periapsis.

In the following paragraphs we show that the dynamical behavior of collision orbits 
changes qualitatively when $\alpha>0$, regardless of how small $\alpha$ is.

This effect is a consequence of the specific structure of the dissipative force. 
The three parameters in equation \eqref{eq:two-bodies-equations} can be reduced 
to the single parameter $c=\frac{C}{\gamma^{2/3}\alpha^{1/3}}$, given in equation 
\eqref{c}, by choosing $\alpha$ as the unit of time and $\gamma^{1/3}\alpha^{2/3}$ 
as the unit of length. With this scaling, system \eqref{eq:two-bodies-equations} becomes
\begin{equation}\label{eq:two-bodies-equations-c}
\begin{cases}
    \dot{r} = v \\
    \dot{v} =\frac{c^2}{r^3} -  \lt(\frac{1}{r^2} +  \frac{v}{r^3}\rt)
\end{cases}
\end{equation}

If $\alpha \ll 1$, the time $\tau$ in equation \eqref{eq:two-bodies-equations-c} is a 
slow time, related to the original time $t$ by $t=\alpha \tau$, while the length scale 
$\lambda$ corresponds to a magnified spatial variable $\ell=\gamma^{1/3}\alpha^{2/3} \lambda$. 
Velocities are correspondingly rescaled as $\frac{\Delta \ell}{\Delta t}=\frac{\gamma^{1/3}}{\alpha^{1/3}}\frac{\Delta \lambda}{\Delta \tau}$.
 
The key point is that, for $\alpha>0$, the limit $C\to 0$ (equivalently $c\to 0$) in 
equation \eqref{eq:two-bodies-equations-c} produces the same effective dynamics as the 
strong dissipative limit obtained by keeping $C>0$ fixed and letting $\alpha\to\infty$. 
Thus, in this model, close approaches between the bodies are intrinsically associated 
with strong energy dissipation.
 
We now return to the original time and length scales used in equation \eqref{eq:two-bodies-equations}. 
For $\alpha>0$ and $C>0$ sufficiently small so that $c=\frac{C}{\gamma^{2/3}\alpha^{1/3}}$ 
is small, most solutions are attracted to the circular periodic orbit, as shown in 
Figure \ref{fig:topologies}. In this regime, the radius $r_c=\frac{C^2}{\gamma}$ is small, 
and the motion rapidly settles into a circular orbit close to the origin. 
 
For $c^6<\frac{1}{4}$, or $c\approx 0.79$, see equation \eqref{eq:eigenvalues-rc},  
$(r_c,0)$ is an attracting node. Thus, $r(t)$ with $r(0)>r_c$ converges to $r_c$ with at 
most one local maximum when $\dot r(0)>0$. 

In contrast, for $\alpha=0$ and fixed small $C>0$, the motion is elliptical with eccentricity 
close to one. If $\alpha>0$ is introduced while keeping $c=\frac{C}{\gamma^{2/3}\alpha^{1/3}}\gg 1$, 
the ellipse slowly shrinks toward the circular orbit of radius $r_c=\frac{C^2}{\gamma}$. 
Increasing $\alpha$ further so that $c<0.79$, or $\alpha > \frac{2C^3}{\gamma^2}$, leads to rapid 
convergence: most of the orbital energy is dissipated near the first perihelion passage, 
and $r(t)\to r_c$ with at most one oscillation. Note that as $C\to 0$, the 
threshold $\alpha > \frac{2C^3}{\gamma^2}$ also tends to zero.

At the same time, as $C\to 0$, the ejection equilibrium $(0, v_e) = \lt(0, \frac{C^2}{\gamma \alpha}\rt)$ 
given in equation \eqref{eq:ejection-orbit} converges to $(0,0)$. Hence, as $C\to 0$, $(0, v_e)$ approaches 
$(r_c,0)$, and these equilibria merge at $C=0$.
 
We now analyze the collision orbits, $C=0$, for any fixed value of $\alpha>0$. 
In this case, equation \eqref{eq:two-bodies-equations-c} becomes
\begin{equation}\label{eq:two-bodies-equations-1}
\begin{cases}
    \dot{r} = v \\
    \dot{v} = -  \lt(\frac{1}{r^2} +  \frac{v}{r^3}\rt)
\end{cases}
\end{equation}
As in equation \eqref{eq:two-bodies-regularized}, we regularize the vector field 
in equation \eqref{eq:two-bodies-equations-1} by multiplying it by $r^3$:
\begin{equation}\label{eq:two-bodies-regularized-1}
\begin{cases}
    \dot{r} = r^3 v \\
    \dot{v} =  -  \lt(r +  v\rt).
\end{cases}
\end{equation}
 
Equation \eqref{eq:two-bodies-regularized-1} has a single equilibrium $(r,v)=(0,0)$, 
with eigenvalues $0$ and $-1$. This equilibrium admits an attracting center 
manifold tangent to the line $v=-r$ and a stable manifold tangent to the $v$-axis. 
A computation shows that the center manifold is given by
\begin{equation}\label{Wc0}
v=h(r)=-r - r^4 (1 + 5 r^3 + 44 r^6 + 539 r^9 + 8337 r^{12} + 154632 r^{15}\ldots )
\end{equation}
Since $h(r)<0$ for $r>0$, equation \eqref{eq:two-bodies-regularized-1} implies that on 
the center manifold $\dot{r}=r^3 h(r)<0$, so $(r,v)=(0,0)$, which represents collisions, 
is attracting.
 
\begin{figure}[H]
    \centering  
    \includegraphics[width=0.5\textwidth]{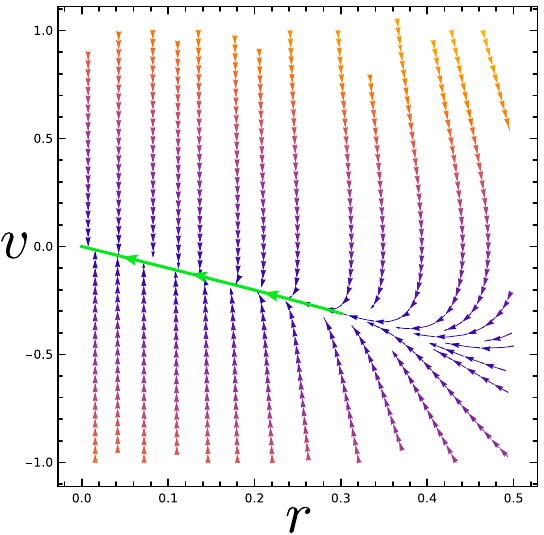}
    \caption{The center manifold of $(r,v)=(0,0)$ as given in equation \eqref{Wc0} and several stream lines of the 
             vector field \eqref{eq:two-bodies-regularized-1}. The topology of the stream lines 
             of equation \eqref{eq:two-bodies-equations-1} is the same.}
    \label{fig:phase}
\end{figure}

The transverse attraction to the center manifold is stronger than the attraction along it; 
thus, solutions rapidly approach the center manifold and then drift along it toward $r=0$. 
Figure \ref{fig:phase} shows the approximation of the center manifold (green) together with 
representative trajectories of \eqref{eq:two-bodies-regularized-1}. 
 
Since the graph $v=h(r)$ is also invariant under the flow of equation \eqref{eq:two-bodies-equations-1}, 
in the dissipative case $\alpha>0$ (with $\alpha$ normalized to one and length scale $\gamma^{1/3}\alpha^{2/3}$ 
equal to one), collisions occur with finite speed approximately given by equation \eqref{Wc0}, as $\dot{r} = v$.

In conclusion, while in the conservative case regularized collision orbits oscillate along a 
segment with infinite velocity at the origin, in the dissipative case, for any $\alpha>0$, a 
collision orbit with $\dot r(0)<0$ moves monotonically toward the origin with finite velocity.

\section{Averaging}\label{sec:averaging}

The goal in this Section is to average equations \eqref{eq:two-bodies-equations} with 
respect to the mean anomaly of the Kepler problem.  We start by writing 
equations \eqref{eq:two-bodies-equations} in an intrinsic way as
\begin{equation}\label{abs}
i_{X}\omega_s=dH+Q
\end{equation}
where $X$ is the dissipative vector field. In the canonical set of variables
$(r,\theta,\dot r, C)$ used in Section \ref{sec:two-bodies}, the symplectic form is 
\[\omega_s=dr\wedge d\dot r +d\theta\wedge dC,\]
 and $Q$, which  we will call  dissipative one-form,
 is given by
\begin{equation}\label{Q}
Q=\alpha\gamma \frac{\dot r}{r^{\beta}} dr=\frac{\alpha\gamma}{2} \frac{\dot r}{r^{\beta+1}} d(r^{2}).
\end{equation}
The angle $\theta$ is the true longitude that can be decomposed as 
\[\theta=\varpi+f
\]
with $f$ as the true anomaly and $\varpi$ as the longitude of the periapsis.

There is a canonical change of variables from $(r,\theta,\dot r, C)$ 
to the Delaunay variables \citep{Laskar2017} given by
\begin{equation}
  \begin{array}{ll}
    M \qquad & \Lambda = \sqrt{\gamma a} \\
    \varpi  \qquad & {\cal H}=C= \Lambda \sqrt{1 - e^2}=\Lambda\eta \\
  \end{array}
\label{eq:Delaunay_vars2d}
\end{equation}
where: $a$ is the semimajor axis, $M$ is the mean anomaly,  and $\Lambda$  is the Delaunay action
conjugated to $M$.
In Delaunay's variables
\[
 \omega_s=dM\wedge d\Lambda+ d\varpi\wedge dC\,, \qquad H=-\frac{\gamma^2}{2\Lambda^2}
\]
It remains to write the dissipative form $Q=\alpha\gamma \frac{\dot r}{r^3} dr$ 
in terms of the Delaunay variables.

In order to write the variables $r$ and $\dot r$ in terms of $M$, $\Lambda$, and the eccentricity $e$
we will use the Hansen's expansions given by:
\begin{equation}
 \left(\frac{r}{a}\right)^{n}\erm^{i m f}=\sum_{k=-\infty}^\infty X^{n,m}_k(e) \erm^{i k M}\,,
 \label{hanseneq}
 \end{equation}
where $n$, $m$, and $k$  are integers and $e$ is the eccentricity.
\footnote{
The Hansen coefficients have the following properties:
\begin{equation}\label{Hanprop}
 X^{n,m}_k\in \mathbb R\,,\qquad 
   X^{n,m}_{-k}= X^{n,-m}_k
 \end{equation}
 and 
 \begin{equation}
   X^{n,m}_k(e)=e^{|m-k|} G^{n,m}_k(e^2)\label{km}
   \end{equation}
where $G^{n,m}_k$ is analytic for $e^2<1$.}

The eccentricity can be written in terms of $\Lambda$ and $C$, and therefore $Q$ can be expressed
in the Delaunay variables.
However, the expression for the dissipative form becomes simpler if we replace $C$ in Delaunay's
variables by the eccentricity $e$. The set of variables $(M,\varpi,\Lambda,e)$ is not 
canonical, so we must compute the symplectic form in this coordinates.

From equation \eqref{eq:Delaunay_vars2d} we obtain
\[
dC= \sqrt{1 - e^2} d\Lambda - \frac{e \Lambda}{\sqrt{1 - e^2}}de =
\eta d\Lambda-\frac{e \Lambda}{\eta}de
\]
In the variables $(M,\varpi, \Lambda, e)$ the symplectic form becomes
\[
 \omega_s=dM \wedge d\Lambda
+\eta \, d\varpi \wedge d\Lambda - \frac{e \Lambda}{\eta}d\varpi\wedge de
\]
As a result the dissipative Hamiltonian equations become:
\begin{equation}\label{Xdis1}
\left(\begin{array}{l}\dot M\\ \dot\varpi\\ \dot\Lambda\\ \dot e\end{array}\right)=\underbrace{\left(\begin{array}{cccc}0&0&1&\frac{\eta^2}{\Lambda e}\\0&0&0&-\frac{\eta}{\Lambda e}\\
-1&0&0&0\\-\frac{\eta^2}{\Lambda e}&\frac{\eta}{\Lambda e}&0&0\end{array}\right)}_{\tbf S}
\left(\begin{array}{c}\partial_M H+Q_M\\\partial_\varpi H+Q_\varpi\\\partial_\Lambda H+Q_\Lambda\\\partial_e H+Q_e\end{array}\right)
\end{equation}
where $Q=Q_MdM+Q_\varpi d\varpi+Q_\Lambda d\Lambda+Q_e de$.

We use the Hansen's expansion to  write\footnote{Consider the Kepler vetor field $X$
given on a subset $U\subset \mathcal R^2\times \mathcal R^2$, for instance, with coordinates $(\tbf x,\tbf v)$,  where all orbits are elliptic. The  elements $a,e,\varpi, C,r,v,f,\ldots$ are 
functions defined on $U$. The vector field $X$ acts on these functions as the time derivative, therefore
\[ X\cdot r=\dot r=v\,,\quad X\cdot e=X\cdot a=0\,,\quad X\cdot f=\dot f\,\ldots\]
The Lie derivative of the differential of any function gives:
$L_Xdr=d (X\cdot r) X=d\dot r$, so the time derivative commutes with the differential. 

In the coordinate system $(M,\Lambda)$, the vector field $X$ is given by $X=n\partial_M$.
We must be cautious in using  the notation $X=n\partial_M$. For instance, if $F=F(r,M)$ 
then $\partial_M F(r,M)$ can be understood as the partial derivative of $F$ with $r=$constant,
which is false, since $\partial_M$ means the derivative of the function while $\Lambda$ is constant.
}
 \[
\frac{\dot r}{r^{\beta+1}}=-\frac{d}{dt}\frac{ r^{-\beta}}{\beta}=
-\frac{1}{\beta}\frac{d}{dt} \left(a^{-\beta}\sum_{j=-\infty}^\infty X^{-\beta,0}_j\erm^{i j M}\right)
\] 
Since  $\dot a=\dot e=0$,  $\dot M=n=\sqrt \gamma a^{-3/2}$,  and $\Lambda=\sqrt{\gamma a}$, we obtain:
\begin{equation}\label{re}
\frac{\dot r}{r^{\beta+1}}=
-i\frac{  \gamma^{\beta+2}}{\beta \Lambda^{2\beta+3}} \sum_{j=-\infty}^\infty j X^{-\beta,0}_j\erm^{i j M}
\end{equation}

We again use Hansen's expansion to write
\begin{equation}\label{dr} \begin{split}
dr^2=&d\left(a^2 \sum_{k=-\infty}^\infty X^{2,0}_k \erm^{i k M}\right)=
d\left(\gamma^{-2}\Lambda^4 \sum_{k=-\infty}^\infty X^{2,0}_k \erm^{i k M}\right)\\
=&i\gamma^{-2}\Lambda^4 \sum_{k=-\infty}^\infty k X^{2,0}_k \erm^{i k M}dM\\
&+4\gamma^{-2}\Lambda^3 \sum_{k=-\infty}^\infty X^{2,0}_k \erm^{i k M}d\Lambda\\
&+\gamma^{-2}\Lambda^4 \sum_{k=-\infty}^\infty \big(X^{2,0}_k\big)^\prime \erm^{i k M}de
\end{split}
\end{equation}
where $\big(X^{2,0}_k\big)^\prime=\frac{d}{de}\big(X^{2,0}_k\big)$.

Therefore, from equations \eqref{Q}, \eqref{re}, and \eqref{dr} we obtain
\begin{equation} \label{dQ}\begin{split}
dQ=&\frac{\alpha \gamma^{\beta+1}}{2\beta \Lambda^{2\beta-1}} \sum_{j=-\infty}^\infty\sum_{k=-\infty}^\infty 
jk X^{-\beta,0}_j   X^{2,0}_k \erm^{i (k+j) M}dM\\
&-2i\frac{\alpha\gamma^{\beta+1}}{\beta \Lambda^{2\beta}} \sum_{j=-\infty}^\infty\sum_{k=-\infty}^\infty j X^{-\beta,0}_j
 X^{2,0}_k \erm^{i (k+j) M}d\Lambda\\
&-i\frac{\alpha \gamma^{\beta+1}}{2\beta \Lambda^{2\beta-1}} \sum_{j=-\infty}^\infty\sum_{k=-\infty}^\infty
 j X^{-\beta,0}_j \big(X^{2,0}_k\big)^\prime \erm^{i (k+j) M}de
\end{split}
\end{equation}

For $\alpha=0$, equations \eqref{Xdis1} become $\dot M=n>0$ and $\dot \varpi=\dot\Lambda=\dot e=0$.
This allows us to average the equations with respect to $M$.  

We define the averaging operator
\begin{equation}\label{avop}
\langle h\rangle=\frac{1}{2\pi}\int_0^{2\pi}h(M)dM
\end{equation}

Since the matrix $\tbf S$ in equation \eqref{Xdis1} does not depend on the mean anomaly the 
averaged equations become 
\begin{equation}\label{av}
\left(\begin{array}{l}\dot M\\ \dot\varpi\\ \dot\Lambda\\ \dot e\end{array}\right)=\left(\begin{array}{cccc}0&0&1&\frac{\eta^2}{\Lambda e}\\0&0&0&-\frac{\eta}{\Lambda e}\\
-1&0&0&0\\-\frac{\eta^2}{\Lambda e}&\frac{\eta}{\Lambda e}&0&0\end{array}\right)
\left(\begin{array}{c}\langle Q_M\rangle\\ \langle Q_\varpi\rangle\\\partial_\Lambda H+\langle Q_\Lambda
\rangle \\ \langle Q_e\rangle\end{array}\right)
\end{equation}

Equation \eqref{dQ} and $X^{n,0}_{-k}=X^{n,0}_{k}$ imply that the averaged components of the dissipative form are:
\begin{equation} \label{dQa}\begin{split}
\langle Q_M\rangle
=&-\frac{\alpha \gamma^{\beta+1}}{2\beta \Lambda^{2\beta-1}} \sum_{k=-\infty}^\infty 
k^2 X^{-\beta,0}_k   X^{2,0}_k \\
\langle Q_\varpi\rangle
=&0\\
\langle Q_\Lambda\rangle
=&2i\frac{\alpha\gamma^{\beta+1}}{\beta \Lambda^{2\beta}} \underbrace{\sum_{k=-\infty}^\infty k X^{-\beta,0}_k
 X^{2,0}_k }_{=0} \\
\langle Q_e\rangle
=&i\frac{\alpha \gamma^{\beta+1}}{2\beta \Lambda^{2\beta-1}}
\underbrace{\sum_{k=-\infty}^\infty
 k X^{-\beta,0}_k \big(X^{2,0}_k\big)^\prime}_{=0} 
\end{split}
\end{equation}

\noindent{\bf Remark}

\begin{equation}
\langle Q_M\rangle=\frac{1}{n}\underbrace{\langle 2\mathcal{D}\rangle}_{\langle\dot E\rangle}=\Delta E \ge 0
\end{equation}
where
\begin{equation}
\Delta E(a,e)=\frac{T}{2\pi}\int_0^{2\pi}\dot E dM=\int_0^{T}\dot E dt
\end{equation}
is the energy dissipated in one period along an elliptic orbit with semi-axis $a$ and eccentricity $e$.

Indeed
\[\begin{split}
\frac{(\tbf v\cdot\hat{\tbf r})^2}{|\tbf r|^\beta}&=\frac{\dot r^2}{r^\beta}=
-\frac{1}{2\beta}\frac{d r^2}{dt}\frac{d r^{-\beta}}{dt}=-\frac{1}{2\beta a^{\beta -2}}\frac{d r^2a^{-2}}{dt}\frac{d r^{-\beta}a^\beta}{dt}\\ &=
-\frac{1}{2\beta a^{\beta -2}}\frac{d }{dt}\left(\sum_jX^{2,0}_j\erm^{ ijM}\right)\frac{d}{dt}\left(\sum_kX^{-\beta,0}_k
\erm^{ ikM}\right)\\ &  =\frac{n^2}{2\beta a^{\beta -2}}
\sum_{j\,,k}X^{2,0}_jX^{-\beta,0}_k j k\erm^{ i(j+k)M}
\end{split}
\]
Thus
\[
\frac{1}{2\pi}\int_0^{2\pi}\frac{(\tbf v\cdot\hat{\tbf r})^2}{|\tbf r|^\beta}dM=
-\frac{n^2}{2\beta a^{\beta -2}}\sum_{k}X^{2,0}_{k}X^{-\beta,0}_k k^2
\]
The result follows from equations \eqref{Q} and \eqref{dQa}.

\begin{figure}[H]
    \centering  
    \includegraphics[width=0.7\textwidth]{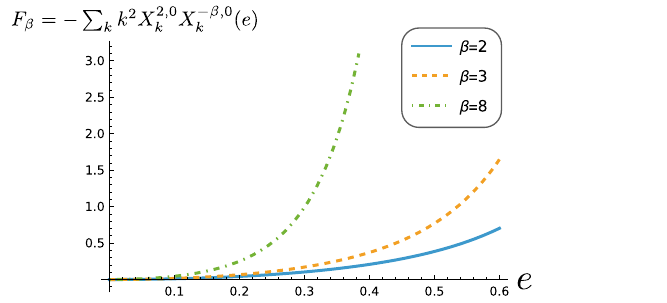}
    \caption{The function $F_\beta(e)=-\sum_k k^2X^{2,0}_k(e)X^{-\beta,0}_k(e)$ for $\beta=2,3,8$.
The following expansions hold: $F_2(e)=e^2+O\left(e^4\right)$,  $F_3(e)=\frac{3 e^2}{2}+O\left(e^4\right)$, and $F_8(e)=4 e^2+O\left(e^4\right)$.}
    \label{fav}
\end{figure}

In Figure \ref{fav} we show three plots of the function $-\sum_k k^2X^{2,0}_k(e)X^{-\beta,0}_k(e)$
for $\beta=2,3,8$. For constant values of $a$ and $e$, this figure shows that the dissipation 
is stronger when $\beta$ increases.

The averaged equations are 
$
\dot M=n$, $\dot\varpi=0$
and 
\begin{equation}
\begin{split}
\dot\Lambda&=-\langle Q_M\rangle<0\\
\dot e&=-\frac{\eta^2}{\Lambda e}\langle Q_M\rangle<0
\end{split}
\label{2bd}
\end{equation}

Using $C=\Lambda\eta$ we obtain that  $C$ remains constant.

Note that $e=0$ is an attracting equilibrium of the equation for $\dot e$. As $e\to 0$, the 
constancy of the angular momentum $C=\Lambda \sqrt{1-e^2}$ implies that $\Lambda\to C$. 
Therefore $a\to C^2/\gamma$ and the asymptotic value of $a$ is determined by the initial angular 
momentum.

The Gauss form of the third Kepler law, namely,
\[
a^3n^2=\frac{C^2}{\tilde p}=m
\]
where $\tilde p=a(1-e^2)=a\eta^2$ is the semi-latus rectum, implies that the semi-latus rectum 
remains constant under the action of the dissipative force.
\vskip .3truecm

One may note that the semi-major axis is always decreasing, so the model cannot explain the 
Earth-Moon increasing distance but it can explain the Mercury-Sun decreasing distance.

\section{Conclusion}\label{sec:conclusions}

% region conclusion
We sought to understand the effect of a dissipative force that while removing energy 
preserves angular momentum. 
The force was motivated by a simplified tidal model 
for punctual masses, where the velocity dependence is linear and projected radially, 
making the force central. 
This form for the force preserves rotational symmetry while breaking time symmetry, 
and is trivially extendable to systems with any number of bodies. 

In particular, we showed that a radial force whose dependence is homogeneous 
of degree $d=3$ yields an extra `radial' symmetry matching that of gravity. 
This allows us to easily derive the equations of motion for planar central 
configurations for generic $N$ bodies, with the resulting system for its 
homographic parameters being equivalent to the dissipative Kepler problem.

% In the derivation of these equations, if one is able to show that the central dissipative force 
% can be expressed as a function of the gravitational potential for a generic degree of 
% homogeneity, the techniques and results obtained here immediately follow for any 
% value of $d$. 

The global phase space portrait for the dissipative Kepler problem was obtained 
via a Poincaré compactification of the regularized vector field. The technique 
used relied on building three different charts to cover the Poincaré disk, where 
the dynamics near infinity can be analyzed locally.

The topology of flow lines in the dissipative Kepler system undergoes a 
transition as the dissipation strength increases, where the separatrix of 
bounded (captured) and unbounded (scattering) orbits in the conservative 
case reconnects with the weak-center-manifold (from $(r_e, v_e)$) linking 
to escaping orbits. 
This reconnection creates a new division between captured and scattered orbits, 
where all approaches between bodies are bounded and lead to circularization.

Collision orbits in the conservative and dissipative settings are fundamentally different. 
While in the conservative case regularized collision orbits oscillate along a segment with 
infinite velocity at the origin, in the dissipative case, for any $\alpha>0$, a collision 
orbit with $\dot r(0)<0$ moves monotonically toward the origin with finite velocity.

As mentioned above, the Kepler system corresponds to the homographic dynamics of 
any planar central configuration under dissipation. 
For sufficiently small dissipation and nonzero angular momentum, 
configurations in the basins of attraction of the circular equilibrium 
have their homothetic (radial) motion damped, leading to purely rotational 
motion and hence to a relative equilibrium. 
For sufficiently strong dissipation, 
any homothethic approach invariably leads to a relative equilibrium, with rotation given by its initial angular momentum.
In the dissipative case, purely homothetic solutions lead to collisions with 
finite velocity, as described above and in more detail in Section \ref{Sec:Coll}.
Although we expect these same topologies for the dissipated Kepler system for any radial 
exponent $d > 1$, the correspondence with homographic motion still relies on the extra symmetry 
of the $d=3$ case.

To extend the current model, one could consider the effect of rotation of bodies, 
reintroducing some of the neglected tidal effects. 
Moreover, to increase flexibility, one may also assign different dissipation strengths 
amongst pairs of bodies, $\alpha \to \alpha_{ij} = \alpha_i \alpha_j$, 
adapting to realistic scenarios where the primary bodies respond differently 
to tides due to its inner structure. 
% endregion

\bmhead{Acknowledgements}

% region acknowledgements
This research was funded by the São Paulo Research Foundation (FAPESP) under 
grants 2024/15052-0 (M. L.) and 2023/07076-4 (C. R.).

C.R. thanks several colleagues with whom he discussed forces similar to the one considered 
in this paper: Claudio Asano, Lucas Ruiz, and especially Antonio Fernandes, who performed several 
computations on the dissipative Kepler problem in a different way that helped in understanding the problem.

Finally, we would like to acknowledge Prof. Mark Levi, who informed us about the PhD thesis of Chong Ai 
and for many interesting conversations on the subject of this paper and related topics.
% endregion

\bmhead{Competing interests}

The authors declare no known competing financial interests or personal relationships that could have influenced the work reported in this paper.

\bibliographystyle{plainnat}
% original: \bibliographystyle{sn-mathphys-ay}
\bibliography{sn-bibliography.bib}

@article{model,
    author  = "First S. Lastname",
    title   = "{C}omprehensive {T}ex {A}rchive {N}etwork ({CTAN})",
    year    = "1993",
    journal = "TUGBoat",
    volume  = "14",
    number  = "3",
    pages   = "342--351",
    DOI     = "10.1002/andp.19053221004"
}

@phdthesis{Chong,
  author  = "Chong Ai",
  title   = "{E}ffect of tidal dissipation on the motion of celestial bodies",
  school  = "The Pennsylvania State University",
  year    = "2012"
}

@article{Breiter1998,
  author  = "S. Breiter and A. A. Jackson",
  title   = "{U}nified analytical solutions to two-body problems with drag",
  journal = "Mon. Not. R. Astron. Soc.",
  volume  = "299",
  year    = "1998",
  pages   = "237--243",
  DOI     = ""
}

@article{Celletti2007,
  author  = "A. Celletti",
  title   = "{W}eakly dissipative systems in Celestial Mechanics",
  journal = "Topics in Gravitational Dynamics. Lecture notes in physics",
  volume  = "729",
  year    = "2007",
  pages   = "67--90",
  DOI     = "10.1007/978-3-540-72984-6_3"
}

@article{Celletti2011,
  author  = "A. Celletti",
  title   = "{S}ome results on the global dynamics of the regularized restricted three-body problem with dissipation",
  journal = "Celestial Mechanics and Dynamical Astronomy",
  volume  = "109",
  year    = "2011",
  pages   = "265--284",
  DOI     = "10.1007/s10569-010-9326-y"
}

@article{Darwin,
  author  = "G. H. Darwin",
  journal = "Philosophical transactions of the Royal Society",
  title   = "{O}n the bodily tides of viscous and semi-elastic spheroids, and on the ocean tides upon a yielding nucleus",
  volume  = "170",
  year    = "1879",
  DOI     = "10.1098/rstl.1879.0061"
}

@article{Diacu1999,
  author  = "F. Diacu",
  journal = "Celestial Mechanics and Dynamical Astronomy",
  title   = "{T}wo-body problems with drag or thrust: qualitative results",
  volume  = "75",
  pages   = "1--15",
  year    = "1999",
  DOI     = "10.1023/A:1008305723295"
}

@article{Ferraz-Mello,
    author  = "S. Ferraz-Mello and
               C. G. Ragazzo and 
               L. R. dos Santos",
    title   = "{D}issipative forces in celestial mechanics",
    year    = "2015",
    journal = "IMPA, Publica{\c{c}}{\~o}es Matem\'aticas: 30o Col{\'o}quio Brasileiro de Matem\'atica"
}

@article{Gabriel,
  author  = "T. S. J. Gabriel and D. J. Scheeres",
  title   = "{E}nergy dissipation end states of the sphere restricted planar three-body problem with collisional interaction",
  journal = "Month Notices of the Royal Astronomical Society",
  volume  = "463",
  pages   = "749--801",
  year    = "2016",
  DOI     = "10.1093/mnras/stw1968"
}

@article{Goldreich,
  author  = "P. Goldreich and Stanton Peale",
  title   = "{S}pin-orbit coupling in the solar system",
  journal = "The Astronomical Journal",
  volume  = "71",
  number  = "6",
  pages   = "425--437",
  year    = "1966",
  DOI     = "10.1086/109947"
}

@article{Margheri2020,
  author  = "A. Margheri and
             M. Misquero",
  title   = "{A} dissipative Kepler problem with a family of singular drags",
  journal = "Celestial Mechanics and Dynamical Astronomy",
  volume  = "132",
  number  = "17",
  year    = "2020",
  DOI     = "10.1007/s10569-020-9956-7"
}

@article{Margheri,
  author  = "A. Margheri and
             R. Ortega and C. Rebelo",
  title   = "{S}ome analytical results about periodic orbits in the restricted three body problem with dissipation",
  journal = "Celestial Mechanics and Dynamical Astronomy",
  volume  = "113",
  pages   = "279--290",
  year    = "2012",
  DOI     = "10.1007/s10569-012-9415-1"
}

@article{Meyer,
  author  = "A. Meyer and
             D. Scheeres and
             H. Agrusa and
             G. Noiset and
             J. McMahon and
             O. Karatekin and
             M. Hirabayashi and R. Nakano",
  title   = "{E}nergy dissipation in synchronous binary asteroids",
  journal = "Icarus",
  volume  = "391",
  pages   = "115323",
  year    = "2023",
  DOI     = "10.1016/j.icarus.2022.115323"
}

@article{Mignard,
  author  = "F. Mignard",
  title   = "{T}he evolution of the lunar orbit revisited. I",
  journal = "The Moon and the Planets",
  volume  = "20",
  pages   = "301--315",
  year    = "1979",
  DOI     = "10.1007/BF00907581"
}

@article{Moeckel2017,
  author  = "R. Moeckel",
  title   = "{M}inimal energy configurations of gravitationally interacting rigid bodies",
  journal = "Celestial Mechanics and Dynamical Astronomy",
  number  = "1",
  volume  = "128",
  pages   = "3--18",
  year    = "2017",
  DOI     = "10.1007/s10569-016-9743-7"
}

@article{Moeckel1990,
  author  = "R. Moeckel",
  title   = "{O}n central configurations",
  journal = "Mathematische Zeitschrift",
  number  = "205",
  volume  = "1",
  pages   = "499--517",
  year    = "1990",
  DOI     = "10.1007/BF02571259"
}

@article{Laskar2017,
  author  = "J. Laskar",
  title   = "{A}ndoyer construction for {H}ill and {D}elaunay variables",
  journal = "Celestial Mechanics and Dynamical Astronomy",
  volume  = "128",
  number  = "4",
  year    = "2017",
  pages   = "475--482",
  DOI     = "10.1007/s10569-017-9761-0"
}

@article{Scheeres2002,
    author  = "D. J. Scheeres",
    title   = "{S}tability in the full two-body problem",
    year    = "2002",
    journal = "Celestial Mechanics and Dynamical Astronomy",
    volume  = "83",
    pages   = "155--169",
    DOI     = "10.1023/A:1020143116091"
}

@article{Scheeres2012,
    author  = "D. J. Scheeres",
    title   = "{M}inimum energy configurations in the N-body problem and the celestial mechanics of granular systems",
    year    = "2012",
    journal = "Celestial Mechanics and Dynamical Astronomy",
    volume  = "113",
    pages   = "291--320",
    DOI     = "10.1007/s10569-012-9416-0"
}

@article{Smale,
    author  = "S. Smale",
    title   = "{T}opology and Mechanics. I",
    year    = "1970",
    journal = "Inventiones mathematicae",
    volume  = "10",
    pages   = "305--331",
    DOI     = "10.1007/BF01418778"
}

@article{Wang,
    author  = "S. Wang and
               G. Huang and X. Wu",
    title   = "{S}imulations of dissipative circular restricted three-body problems using the velocity-scaling correction method",
    year    = "2018",
    journal = "The Astronomical Journal",
    volume  = "155",
    number  = "67",
    pages   = "1--12",
    DOI     = "10.3847/1538-3881/aa9ff9"
}

@book{Contopoulosbook,
    author    = "G. Contopoulos",
    title     = "Order and Chaos in Dynamical Astronomy",
    year      = "2002",
    publisher = "Springer",
    address   = "Berlin, Heidelberg"
}

@book{Dumortierbook,
  author    = "F. Dumortier and J. Llibre and J. C. Art{\\'e}s",
    title     = "Qualitative theory of Planar Differential Systems",
    year      = "2006",
    publisher = "Springer",
    address   = "Berlin"
}

@book{Ferraz-Mellobook,
    author    = "S. Ferraz-Mello",
    title     = "Caos e Planetas: din\^{a}mica ca\'{o}tica de sistemas planet\'{a}rios",
    year      = "2021",
    publisher = "Editora Livraria da F\'{i}sica",
    address   = "S\~{a}o Paulo"
}

@book{Moeckelbook,
    author    = "J. Llibre and R. Moeckel and C. Sim\'{o}",
    title     = "Central configurations, periodic orbits, and Hamiltonian systems",
    year      = "2015",
    publisher = "Springer",
    address   = "Basel"
}

\vspace{1.0cm}
\centering \pgfornament[width = 8cm, color = black]{89}

\end{document}